\documentclass[pre,twocolumn,a4paper,superscriptaddress,showpacs,showkeys]{revtex4}
\pdfoutput=1
\usepackage[english]{babel}
\usepackage{amsmath}
\usepackage{amsfonts}
\usepackage{amssymb}
\usepackage{array}
\usepackage{dcolumn}
\usepackage{longtable}
\usepackage{hyperref}
\usepackage{float}
\usepackage{bbm}
\usepackage{bm}
\usepackage{graphicx}
\usepackage{natbib}

\def\d{\text{d}}

\def\dt{\partial_t}
\def\h{_\text{h}}

\def\uzt{u_{z\theta}}

\begin{document}
\title{Mechanochemical action of the dynamin protein}
\date{\today}
\author{Martin Lenz}
\email{martin.lenz@curie.fr}
\affiliation{Institut Curie, Centre de Recherche, Laboratoire Physico-Chimie Curie, Paris, F-75248 France; CNRS, UMR 168, Paris, F-75248 France; Universit\'e Pierre et Marie Curie-Paris6, UMR 168, Paris, F-75005 France}
\author{Jacques Prost}
\affiliation{Institut Curie, Centre de Recherche, Laboratoire Physico-Chimie Curie, Paris, F-75248 France; CNRS, UMR 168, Paris, F-75248 France; Universit\'e Pierre et Marie Curie-Paris6, UMR 168, Paris, F-75005 France}
\affiliation{E.S.P.C.I., 10 rue Vauquelin, F-75231 Paris Cedex 05, France}
\author{Jean-Fran\c{c}ois Joanny}
\affiliation{Institut Curie, Centre de Recherche, Laboratoire Physico-Chimie Curie, Paris, F-75248 France; CNRS, UMR 168, Paris, F-75248 France; Universit\'e Pierre et Marie Curie-Paris6, UMR 168, Paris, F-75005 France}

\begin{abstract}
Dynamin is a ubiquitous GTPase that tubulates lipid bilayers and is implicated in many membrane severing processes in eukaryotic cells. Setting the grounds for a better understanding of this biological function, we develop a generalized hydrodynamics description of the conformational change of large dynamin-membrane tubes taking into account GTP consumption as a free energy source. On observable time scales, dissipation is dominated by an effective dynamin/membrane friction and the deformation field of the tube has a simple diffusive behavior, which could be tested experimentally. A more involved, semi-microscopic model yields complete predictions for the dynamics of the tube and possibly accounts for contradictory experimental results concerning its change of conformation as well as for plectonemic supercoiling.
\end{abstract}


\maketitle

\section{\label{sec:introduction}Introduction}
In order to operate efficiently, living cells must constantly 
maintain concentration gradients of various chemical species and 
isolate some of their components. One of the many different 
biological processes required to maintain this traffic 
is membrane fission, by which a cell membrane compartment is 
split into two or more topologically distinct parts. A fundamental 
protein involved in most membrane fission events is dynamin, 
which has been proposed to be a ``universal membrane 
fission protein''~\cite{Praefcke:2004aa}. Dynamin and its 
analogues are found in cellular processes as diverse 
as clathrin-coated endocytosis, phagocytosis, mitochondria 
and chloroplasts reorganization, cell division and virus 
immunization in organisms ranging from mammals to yeast 
and plants \cite{Danino:2001aa, Praefcke:2004aa}. In most of 
these processes, two separating membrane compartments 
end up being linked by a thin membrane neck which 
is difficult to sever, since the membrane must be strongly curved before it can be pinched off. Dynamin-like proteins localize at this neck and participate 
in its breaking, thus completing the fission. 
Although mutations in such an important protein are often lethal, 
defective dynamin or dynamin analogues have been shown to be involved 
in human diseases such as the optical atrophy type 1, the 
Charcot-Marie-Tooth disease and the dominant centronuclear 
myopathy \cite{Alexander:2000aa, Zuchner:2005aa,Bitoun:2005aa}.

The role of dynamin in tube fission was first suggested by results 
showing the importance of its \emph{Drosophila} analogue 
in endocytosis \cite{Koenig:1989aa}. Dynamin is recruited by clathrin-coated vesicles, possibly through membrane-mediated elastic interactions \cite{Fournier:2003aa}, and self-assembles into short (a few helical repeats) helical constructs on the cell membrane necks localized at their base \cite{Takei:1995aa}. Much longer (thousands of helical repeats) helical dynamin 
polymers have been observed in cell-free environments, 
either wrapped around microtubule templates \cite{Shpetner:1989aa}, or 
in solution \cite{Hinshaw:1995aa, Carr:1997aa}. Purified dynamin 
also polymerizes around negatively-charged lipid 
bilayers \cite{Takei:1998aa,Sweitzer:1998aa}, deforming 
liposomes into dynamin-coated nanotubes, simply termed ``tubes'' in 
the following. Electron micrographs suggest that these tubes are hollow, 
i.e. filled with water \cite{Zhang:2001aa,Chen:2004aa,Mears:2007aa}. 
Dynamin is a GTPase and therefore catalyses the hydrolysis of guanosine 
triphosphate (GTP) into guanosine diphosphate (GDP) and inorganic 
phosphate (Pi). This highly exoenergetic reaction ($\sim 25\,k_BT$ per GTP molecule in a 
typical cellular environment) is similar to the hydrolysis of adenosine triphosphate (ATP), 
which fuels most known molecular motors and many other cellular processes \cite{Alberts:1998aa}, and has 
been shown to be necessary for endocytosis \cite{Carter:1993aa}. Self-assembly 
of dynamin has been linked to a dramatic increase of its GTPase 
activity \cite{Shpetner:1992aa,Maeda:1992aa,Tuma:1993aa}, thus suggesting 
that GTP hydrolysis by self-assembled dynamin drives membrane tube 
fission during endocytosis \cite{Sever:2000aa, Danino:2001aa, Praefcke:2004aa}.

During the past decade, experimental evidence indicating that 
tube breaking involves a mechanochemical action of dynamin 
has been accumulated. Electron microscopy has shown that the geometry of the
dynamin helical coat changes upon GTP hydrolysis. However, 
there is still a controversy as to whether both the pitch 
and radius of the helix shrink
\cite{Sweitzer:1998aa,Zhang:2001aa,Danino:2004aa,Chen:2004aa,Mears:2007aa} 
(see Fig.~\ref{fig:dynamin_symmetry}(a)) or the pitch increases 
at constant radius \cite{Stowell:1999aa,Marks:2001aa}. In the absence 
of any other protein than dynamin, this change of conformation 
is sufficient to drive tube breakage when the end points of the tube 
are attached to a substrate \cite{Sweitzer:1998aa}. However, 
no fission of freely floating tubes is observed \cite{Danino:2004aa}. 
More recently, optical microscopy has been used to investigate 
the dynamics of the tube's conformational change and breaking \cite{Roux:2006aa}. In these experiments, dynamin-coated 
membrane nanotubes are grown from a lamellar phase of a suitable 
lipid mixture. Then GTP is injected in the experimental chamber 
(typically in a few tenths of seconds). 
The initially rather floppy tubes then straighten, 
revealing a build up of their tension. 
If one end of the tube is free to fluctuate, this tension results in 
the retraction of the tube. If both ends are attached, the tube breaks. 
If polystyrene beads (diameter 260-320\,nm) are attached to 
the dynamin coat, rotation is observed after GTP injection, 
showing that GTP hydrolysis induces not only tension 
but also torques in the tubes. The typical time scales involved in the rotation of the beads and the breaking of the tubes 
are roughly 3\,s after GTP injection.

In this article, we present a theoretical model for
the relaxation of long dynamin-coated membrane nanotubes accounting for the above-mentioned experimental results. We believe that 
a quantitative description of the tube dynamics will help to understand 
the mechanism by which dynamin severs membrane tubes. This is a 
much-debated question for which several models have been 
proposed \cite{Sever:2000aa}. Since little quantitative information 
about the microscopic details of the dynamin helix is available, we choose 
a coarse-grained (hydrodynamic) approach. In this framework, 
we do not need to speculate about the unknown microscopic details of 
the non-equilibrium behavior of the tube: its dynamics is characterized by a few phenomenological
transport coefficients.

In the next three sections, we present the building blocks of our formalism 
by decreasing order of generality. In Sec.~\ref{sec:Hydrodynamic theory}, 
we consider only the symmetries of the system and write the most 
general hydrodynamic theory compatible with these symmetries. 
We then argue in Sec.~\ref{sec:Long times dynamics for the 
helix/membrane friction-limited regime} that one hydrodynamic mode 
is much slower than the others. This leads to simplified 
equations describing this mode. In Sec.~\ref{sec:Susceptibility matrices describing experimental situations}, we present two microscopic models of the equilibrium properties of the 
tube aimed at describing two possible experimental situations. This allows us to solve the equations of motion and make predictions about the tube dynamics. In Sec.~\ref{sec:Comparison to experimental results}, we compare these predictions to experimental results and thus justify some of our assumptions. A tentative account of the differences in the conformational changes of dynamin reported in Refs.~\cite{Sweitzer:1998aa,Zhang:2001aa,Danino:2004aa,Chen:2004aa,Mears:2007aa} on the one hand and Refs.~\cite{Stowell:1999aa,Marks:2001aa} on the other hand is also given. Finally, we discuss the generality of our model and its implications for membrane nanotube fission in Sec.~\ref{sec:Discussion}.

\section{\label{sec:Hydrodynamic theory}Hydrodynamic theory}
\begin{figure}
\resizebox{8.5cm}{!}{\includegraphics{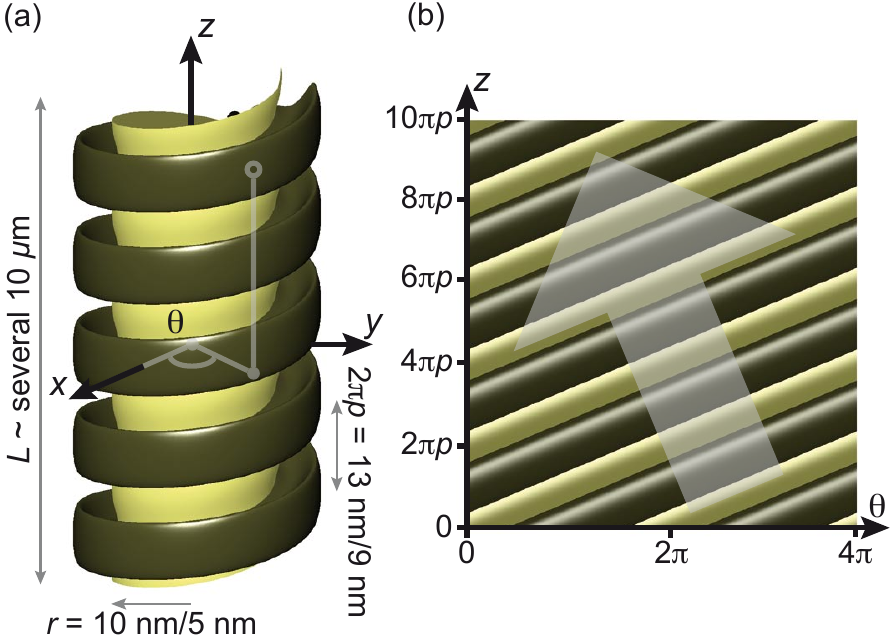}}
\caption{\label{fig:dynamin_symmetry}Schematics representing the geometry of the tube. (a)~The tube comprises two fluids $h$ 
and $m$, here pictured in different colors. It is invariant under a 
rotation around $z$ by an arbitrary angle $\theta$ followed by a translation
by $p\theta$ along $z$ (we refer to this property as \emph{helical symmetry 
with pitch }$2\pi p$ in the text). Moreover, the system is assumed to 
be invariant under a rotation of $\pi$ around the $x$-axis (the system 
is \emph{non-polar}). The latter transformation is equivalent to a reversal 
of polar coordinates $(\theta,z)\rightarrow (-\theta,-z)$. The numerical 
values measured in Ref.~\cite{Chen:2004aa} for the radii and pitches in 
the relaxed/constricted states are indicated on the figure. 
(b)~Representation of the same helix in the $\theta$, $z$ plane of 
cylindrical coordinates with periodic boundary conditions on $\theta$. 
The system clearly has a broken translational symmetry in the direction of
the translucent arrow. It is described by a broken-symmetry variable $\uzt$ obeying Eq.~(\ref{eq:uzt_conservation}), which can be understood as a conservation law for the number of stripes visible here. The associated reactive current is $v_h/p-\Omega_h$, the projection of the velocity of the stripes on the direction of the arrow.}
\end{figure}

In this section, we derive equations of motion for dynamin/membrane tubes 
based on the symmetries of the system. We also restrict our study to the long length 
and time scales, thus constructing a \emph{hydrodynamic theory}. More specifically, we focus on the so-called \emph{hydrodynamic modes}, which are spatially inhomogeneous excitations of the system away from equilibrium with the following properties \cite{Martin:1972aa}: 
\begin{enumerate}
\item the amplitude of these excitations are small enough for the system to 
remain weakly out of equilibrium in the sense of Ref.~\cite{DeGroot:1984aa},
\item the wave vector $q$ and the pulsation $\omega(q)$ 
characterizing the spatial inhomogeneity of 
the hydrodynamic mode are such that
\begin{equation}\label{eq:hydrodynamic}
\lim_{q\to 0} \omega(q)= 0.
\end{equation}
\end{enumerate}

The $q\to 0$ limit corresponds to excitations over length scales much 
larger than the microscopic length scales of the system. Typically, we consider 
inverse wave vectors of the order of the tube's length: $q^{-1}\sim \text{several } 10\,\mu$m. This is indeed much larger than the typical microscopic length: the tube radius $r\sim 10\,$nm. 
The hydrodynamic theory thus involves a coarse-graining of the system at 
the scale of a few tens of nanometers, and so the tube must be treated as 
a one-dimensional object.

Let us now state the hypotheses underlying our hydrodynamic theory. We consider 
a one-dimensional system comprising two fluids which we refer to as 
fluids $h$ (representing the helix) and $m$ (the lipid membrane). In agreement with electron microscopy data \cite{Zhang:2001aa,Chen:2004aa,Mears:2007aa}, 
we assume that the system has a helical symmetry with pitch $2\pi p$ and is non-polar, as shown on Fig.~\ref{fig:dynamin_symmetry}(a).

In the following, we identify the relevant variables describing this system and derive an expression for its entropy production. Introducing an active term representing the input of free energy in the form of GTP, we write the constitutive (flux/force) equations for the tube. Together with conservation laws these equations eventually yield the hydrodynamic modes of the system.

\subsection{\label{sec:Conservation laws and hydrodynamic variables}Conservation laws and hydrodynamic variables}
The first step in building a hydrodynamic theory relies on conservation laws.

We assume that no exchange of membrane or dynamin occur with the aqueous medium surrounding the tube on the time scale of the change of conformation \cite{Roux:2006aa}. Therefore, the masses of fluids $h$ and $m$ obey the following conservation equations:
\begin{subequations}\label{eq:rho_hm_conservation}
\begin{eqnarray}
\dt \rho_h&=&-\nabla (\rho_hv_h)\label{eq:rho_h_conservation}\\
\dt \rho_m&=&-\nabla (\rho_mv_m),\label{eq:rho_m_conservation}
\end{eqnarray}
\end{subequations}
where $\nabla$ is the differentiation operator with respect to $z$. $v_h$ and $v_m$ are the velocities of fluids $h$ and $m$ respectively and $\rho_h$ and $\rho_m$ their mass densities (masses per unit of $z$ length). We now define the mass fraction of $h$ as $\Phi=\rho_h/\rho$, the mass density of the whole tube $\rho=\rho_h+\rho_m$, the linear momentum density of the tube $g=\rho_hv_h+\rho_mv_m=\rho v$ with $v$ the center-of-mass velocity and the diffusion flux of $h$ relative to the center of mass of the tube $J=\rho_h(v_h-v)$. The conservation laws expressed in Eqs.~(\ref{eq:rho_hm_conservation}) can be re-written as
\begin{subequations}\label{eq:rho_Phi_conservation}
\begin{eqnarray}
\dt \rho&=&-\nabla g\label{eq:rho_conservation}\\
\dt \Phi&=&-v\nabla\Phi-\rho^{-1}\nabla J,\label{eq:Phi_conservation}
\end{eqnarray}
\end{subequations}
It is shown further below that the inverse relaxation times of $\rho$ and $\Phi$ go to zero with vanishing $q$, meaning that $\rho$ and $\Phi$ are hydrodynamic variables.

Let $l$ be the angular momentum density of the tube. The conservation laws for $g$ and $l$ are the force and torque balance equations. There are two contributions to the force (resp. torque) applied to a tube element: first, the divergence of $\sigma$ (resp. $\tau$), the linear (resp. angular) internal stress of the tube; and second, the external force (resp. torque) due to the coupling of the helix dynamics with the hydrodynamic flow that it induces in the surrounding aqueous medium. For simplicity, we model this ``friction against water'' as a force and a torque linearly dependent of $v_h$ and the angular velocity $\Omega_h$ of fluid $h$ with proportionality coefficients $\lbrace\gamma_{ij}\rbrace_{i,j=z,\theta}$:
\begin{subequations}\label{eq:gl_conservation}
\begin{eqnarray}
\dt g&=&-\nabla\sigma -\gamma_{zz}v_h-\gamma_{z\theta}\Omega_h\\
\dt l&=&-\nabla\tau -\gamma_{\theta z}v_h-\gamma_{\theta\theta}\Omega_h.
\end{eqnarray}
\end{subequations}
Since $v_h=(g+J/\Phi)/\rho$, we can replace $v_h$ by $g+J/\Phi$ in Eqs.~(\ref{eq:gl_conservation}) after rescaling the friction coefficients in the following way: $\gamma_{zz}\rightarrow\rho\gamma_{zz}$ and $\gamma_{\theta z}\rightarrow\rho\gamma_{\theta z}$. Since the friction between the two fluids is a local phenomenon (which does not depend on the wave vector $q$), the quantity $\Omega_h-\Omega_m$ relaxes to zero in a non-hydrodynamic time (that does not respect Eq.~(\ref{eq:hydrodynamic})). When studying hydrodynamic time scales, we can thus replace $\Omega_h$ by $\Omega_m$ or equivalently by $\Omega=l/I$ with $I$ the tube's density of moment of inertia. Redefining $\gamma_{z\theta}\rightarrow I\gamma_{z\theta}$ and $\gamma_{\theta\theta}\rightarrow I\gamma_{\theta\theta}$, we replace $\Omega_h$ by $l$ in Eqs.~(\ref{eq:gl_conservation}).

It is now apparent in Eqs.~(\ref{eq:gl_conservation}) that in the presence of friction against water, $g$ and $l$ relax to the solutions of the following equations:
\begin{subequations}\label{eq:gl_overdamped}
\begin{eqnarray}
\gamma_{zz}\left(g+\frac{J}{\Phi} \right)+\gamma_{z\theta}l&=&-\nabla\sigma\\
\gamma_{\theta z}\left(g+\frac{J}{\Phi} \right)+\gamma_{\theta\theta}l&=&-\nabla\tau
\end{eqnarray}
\end{subequations}
over times of order $1/\gamma_{ij}$. Since the $\gamma_{ij}$s do not depend on $q$, these times are not hydrodynamic times (they do not go to infinity when $q$ vanishes). The linear and angular momentum densities $g$ and $l$ are therefore not hydrodynamic variables and are given by Eqs.~(\ref{eq:gl_overdamped}), which are the force and torque balance equations in the overdamped regime.

The conservation of energy reads:
\begin{eqnarray}
\dt\varepsilon+\nabla j^\varepsilon&=&v\left(-\gamma_{zz} \left(g+\frac{J}{\Phi} \right)-\gamma_{z\theta} l\right)\nonumber\\
&&+\Omega\left(-\gamma_{\theta z} \left(g+\frac{J}{\Phi} \right)-\gamma_{\theta\theta} l\right),\label{eq:energy_conservation}
\end{eqnarray}
where $\varepsilon$ is the energy density and $j^\varepsilon$ the current of energy. The right-hand side accounts for energy dissipation by friction.

Although no other conservation law than Eqs.~(\ref{eq:rho_Phi_conservation}), (\ref{eq:gl_overdamped}) and (\ref{eq:energy_conservation}) exist in the system, its hydrodynamic description is still incomplete. Indeed, the system has a broken continuous symmetry similar to that of smectic-A liquid crystal phases (Fig.~\ref{fig:dynamin_symmetry}(b)). Just as in the case of liquid crystals~\cite{deGennes:1993aa}, we use a strain tensor component $\uzt$ to describe this symmetry breaking. We define it in the following way: let $\theta(z,t)$ be the angular displacement of the intersection of the helix with the plane located at altitude $z$ at time $t$ (therefore $\theta$ is a eulerian coordinate). The strain is then defined as:
\begin{equation}
\uzt(z,t)=\frac{\partial \theta}{\partial z}(z,t).
\end{equation}
As any broken-symmetry variable, $\uzt$ obeys a relation similar to a conservation law. To show this, we first note that in a fully reversible situation
\begin{equation}
\dt \theta (z,t)=\Omega_h-\frac{v_h}{p}.
\end{equation}
Differentiating this equation with respect to $z$, one finds
\begin{equation}
\dt \uzt= -\nabla\left(\frac{v_h}{p}-\Omega_h\right),
\end{equation}
where $v_h/p-\Omega_h$ is the reactive current of $\uzt$ (see also Fig.~\ref{fig:dynamin_symmetry}(b)). In the presence of dissipation, one must add a dissipative part $X$ to this current \cite{Chaikin:1995aa}, hence
\begin{eqnarray}
\dt \uzt&=& -\nabla X-\nabla\left(\frac{v_h}{p}-\Omega_h\right)\nonumber\\
&=&-\nabla\left(X+\frac{g+J/\Phi}{\rho p}-\frac{l}{I}\right).\label{eq:uzt_conservation}
\end{eqnarray}

\subsection{Entropy production}
The great simplicity of hydrodynamic theories can be tracked back to the fact that in the long-time limit, one locally describes the state of a potentially very complex system using only a few conserved quantities. Indeed, on time scales going to infinity with the size of the system, all the microscopic (fast) degrees of freedom have relaxed and the system is locally in a state of thermal equilibrium in the thermodynamic ensemble defined by the conserved quantities.

Let us apply this idea to the tube in the general case where friction against water is not necessarily present (i.e. the $\gamma_{ij}$s can be zero, in which case $g$ and $l$ are hydrodynamic variables). The state of local thermal equilibrium is entirely characterized by the six quantities 
$g$, $l$, $\rho$, $\Phi$, $\uzt$ and $\varepsilon$. Equivalently, we can consider a homogeneous tube of length $V$ and study 
it in the thermodynamic ensemble $(P, L, M,\Phi,\uzt,T,V)$, where $P=Vg$, $L=Vl$ 
and $M=V\rho$ are the total linear momentum, angular momentum and mass 
of the tube; $T$ is the temperature. The total differential of the free energy of the tube reads:
\begin{equation}\label{eq:free_energy_differential}
\d F=v\d P+\Omega\d L+\mu\d M+M\mu_e\d \Phi+H\d\uzt-S\d T-\mathfrak{p}\d V.
\end{equation}
This equation defines the total and exchange chemical potentials $\mu$ and $\mu_e$, the reactive stress $h=H/V$, the entropy $S=Vs$ and the equilibrium pressure $\mathfrak{p}$. Note that the velocity $v$ is thermodynamically conjugated to $P$ and the angular velocity $\Omega$ to $L$. This implicitly assumes that the free energy is 
the sum of a kinetic energy and a static contribution:
\begin{equation}\label{eq:F=F0+Ec}
F=V\left(\frac{\rho v^2}{2}+\frac{I\Omega^2}{2}\right)+F_0(M,\Phi,\uzt,T,V),
\end{equation}
where the free energy of the tube in its rest frame can also be written as
\begin{equation}\label{eq:f0}
F_0(M,\Phi,\uzt,T,V)=Vf_0(\rho,\Phi,\uzt,T).
\end{equation}
Using an extensivity argument, one shows that
\begin{equation}\label{eq:pressure}
\mathfrak{p}=vg+\Omega l+\mu\rho+Ts-\varepsilon
\end{equation}
and proves the local form of the fundamental equilibrium thermodynamic relation:
\begin{equation}\label{eq:entropy_differential}
T\d s=-v\d g-\Omega\d l-\mu\d\rho-\rho\mu_e\d\Phi-h\d\uzt+\d\varepsilon.
\end{equation}

Inserting the conservation equations Eqs.~(\ref{eq:rho_Phi_conservation}), (\ref{eq:gl_conservation}), (\ref{eq:energy_conservation}) 
and (\ref{eq:uzt_conservation}) into (\ref{eq:entropy_differential}) and 
using (\ref{eq:pressure}), one finds the following form for the local 
entropy production of the system \cite{Chaikin:1995aa}:
\begin{eqnarray}\label{eq:entropy_production}
T\left(\frac{\partial s}{\partial t}
+\nabla\left(vs+\frac{Q}{T}\right)\right)=&-&\left(\sigma-\mathfrak{p}-
\frac{h}{p}\right)\nabla v\nonumber\\
&-&(\tau+h)\nabla\Omega\nonumber\\
&-&J\nabla\mu_e\nonumber\nonumber\\
&-&\left(X+\frac{J}{\rho\Phi p}\right)\nabla h\nonumber\\
&-&Q\frac{\nabla T}{T},
\end{eqnarray}
where $Q$ is the heat current in the $z$ direction and where higher-order terms in the displacement from equilibrium have been dropped.

\subsection{\label{sec:Constitutive equations}Constitutive equations}
The right-hand side of Eq.~(\ref{eq:entropy_production}) is the sum 
of five terms, each of which is the product of a \emph{flux} and a \emph{force} 
as displayed in Table~\ref{tab:fluxes_and_forces}. These fluxes and forces 
vanish at thermal equilibrium. Also, according to the second law 
of thermodynamics, entropy production is always positive. 
Therefore, close to the equilibrium state, the fluxes depend linearly 
on the forces through a positive definite matrix. In addition to this positivity condition, other constraints exist on 
the relationships between fluxes and forces:

\begin{table}
\caption{\label{tab:fluxes_and_forces}
The fluxes, forces and signature of the forces under two symmetry operations: ``time symmetry'' denotes the time-reversal symmetry and ``spatial symmetry'' refers to the reversal of the polar coordinates $(\theta,z)\rightarrow (-\theta,-z)$ defined in Fig.~\ref{fig:dynamin_symmetry}(a).}
\begin{ruledtabular}
\begin{tabular}{cccc}
Flux&Force&time symmetry&spatial symmetry\\
\hline
$\sigma-\mathfrak{p}-h/p$ & $\nabla v$ & - & +\\
$\tau+h$ & $\nabla\Omega$ & - & +\\
$J$ & $\nabla\mu_e$ & + & -\\
$X+J/(\rho\Phi p)$ & $\nabla h$ & + & -\\
$Q$ & $\nabla T/T$ & + & -\\
 & $\Delta\mu$ & + & +
\end{tabular}
\end{ruledtabular}
\end{table}

First, entropy production is invariant under spatial symmetry operations that leave the system unchanged. Therefore, fluxes and forces of opposite signature under the spatial symmetry $(\theta,z)\rightarrow (-\theta,-z)$ defined in Fig.~\ref{fig:dynamin_symmetry}(a) cannot be coupled. This property is a special case of the Curie principle, which states that in an isotropic system, couplings between fluxes and forces of different tensorial characters are forbidden. In this context, the quantities displayed in Table~\ref{tab:fluxes_and_forces} that are odd under the transformation $(\theta,z)\rightarrow (-\theta,-z)$ are analogous to vectors and those that are even are scalars or second-rank tensors \cite{DeGroot:1984aa}.

Time-reversal symmetry imposes another
set of constraints. Each flux can be written as the sum of a dissipative 
part, which has the same time symmetry as the conjugate force, and a 
reactive part with the opposite symmetry. \emph{Dissipative couplings} 
occur between fluxes and forces having the same time-reversal symmetry. Conversely, \emph{reactive couplings} relate fluxes and forces of opposite symmetries. According to Onsager's relations, the matrix 
of dissipative couplings is symmetric and the matrix of reactive 
couplings antisymmetric \cite{Landau:1980aa}.

While deriving Eq.~(\ref{eq:entropy_production}), we have made sure 
that its right-hand side involves only entropy production terms, and 
no entropy exchange or energetic effects. Therefore, the fluxes in 
this equation are dissipative and have a vanishing reactive part. 
Taking into account the symmetry constraints discussed above, we obtain
the following set of constitutive equations:
\begin{subequations}\label{eq:constitutive_1}
\begin{eqnarray}
\sigma-\mathfrak{p}-h/p &=& -\eta_z \nabla v - a\nabla\Omega,\label{eq:constitutive_1_1}\\
\tau+h &=& -a \nabla v - \eta_\theta\nabla\Omega,\label{eq:constitutive_1_2}\\
J &=& - \lambda\nabla\mu_e - b\nabla h - c\nabla T/T,\label{eq:constitutive_1_3}\\
X+J/(\rho\Phi p) &=& - b\nabla\mu_e - \tilde{\lambda}\nabla h - d\nabla T/T,\label{eq:constitutive_1_4}\\
Q &=& - c\nabla\mu_e - d\nabla h - \kappa\nabla T/T.\label{eq:constitutive_1_5}
\end{eqnarray}
\end{subequations}
The coefficients in front of the forces are so-called \emph{phenomenological transport coefficients}. They are \emph{a priori} unknown coefficients that depend on the microscopic details of the problem.

\subsection{\label{sec:Discussion of the phenomenological coefficients}Discussion of the phenomenological coefficients}
In the spirit of the present article, those phenomenological coefficients 
that are relevant to the relaxation of the system should be 
determined experimentally. The only way to calculate them 
\emph{a priori} would be to use a detailed microscopic model, which would 
require a better knowledge of dynamin than we have. However, in the next 
few paragraphs, we try to interpret the origin and give typical orders 
of magnitude of these phenomenological coefficients.

The coefficient $\eta_z>0$ can be identified as 
a $\text{length}\times\text{surface viscosity}$, where ``length'' denotes  
a typical microscopic length of the tube, for instance its inner 
radius $r\sim 10\,$nm. Similarly, $\eta_\theta>0$ is a 
$(\text{length})^3\times\text{surface viscosity}$. Assuming that the effective 
characteristic surface viscosity of the tube is close to that of a 
lipid bilayer, namely of order $5\times 10^{-9}\,\text{kg.s}^{-1}$ 
\cite{Waugh:1982aa}, we estimate that $\eta_z\sim 10^{-16}\,
\text{kg.m.s}^{-1}$ and $\eta_\theta\sim 10^{-32}\,\text{kg.m}^3
\text{.s}^{-1}$.

The momentum transfer from translational to rotational 
degrees of freedom is described by $a$. This transfer is allowed since the tube is chiral. 
The amplitude of these effects is constrained by the positivity of 
the matrix of phenomenological coefficients, which imposes $\vert a\vert<\sqrt{\eta_z\eta_\theta}$.

The coefficient $\lambda>0$ relates a gradient of chemical potential to a diffusion flux. 
By analogy to Fick's law, we expect it to be proportional to 
a diffusion coefficient. To better interpret $\lambda$, let us set all 
other phenomenological coefficients to zero. The hydrodynamic dissipation then comes only from the homogenization of the helix mass fraction $\Phi$ 
at fixed mass density $\rho$, and therefore involves a relative flow between 
the two fluids $h$ and $m$. In this scenario, the source of dissipation 
is obviously the friction between the two fluids. One can therefore interpret $\lambda$ as the inverse of a helix/membrane friction coefficient. 
In order to calculate an order of magnitude, we propose an 
oversimplified friction mechanism inspired by Ref.~\cite{Burger:2000aa}, where it is shown that dynamin inserts into the outer leaflet of the membrane bilayer. In this naive model, we assume that the outer membrane 
monolayer is attached to the helix and that the energy dissipation
comes from the sliding of one monolayer against the other. Experimentally, one measures typical friction coefficients for the relative sliding of lipid monolayers of order $\beta=10^8\,$Pa/(m.s$^{-1}$) \cite{Evans:1994aa}. Let us consider a motionless isothermal ($\nabla T=0$) cylinder of membrane of length $L$ surrounded by an undeformed ($h=0$) helix of dynamin moving at velocity $v_h$ under the influence of a chemical potential gradient difference $\mu_e=L\nabla\mu_e$ between the extremities of the cylinder. The mass flow of helix in such a system is $\rho_hv_h=\rho\Phi v_h$, hence the tube receives a net power ${\cal P}=\mu_e\rho\Phi v_h$ from the reservoirs located at each end of the cylinder. Eq.~(\ref{eq:constitutive_1_3}) and $J=\rho\Phi(1-\Phi)v_h$ entail ${\cal P}/L=\rho^2\Phi^2(1-\Phi)v_h^2/\lambda$. Assuming that this power is entirely dissipated by the friction between the membrane and the helix implies ${\cal P}/L=2\pi r\beta v_h^2$ and eventually
\begin{equation}
\lambda=\frac{\Phi^2(1-\Phi)}{2\pi}\frac{\rho^2}{r\beta}\simeq 1.1\times 10^{-26}\,\text{kg.m}^{-1}\text{.s},
\end{equation}
where the values at equilibrium $\rho_{h0}=\rho_0\Phi_0\simeq 3.7\times 10^{-13}\,$kg.m$^{-1}$ and $\rho_{m0}=\rho_0(1-\Phi_0)\simeq 3.8\times 10^{-13}\,$kg.m$^{-1}$ are calculated from the molecular mass of dynamin \cite{Sever:2000aa} and the number of dynamin monomers per helix turn \cite{Chen:2004aa} on the one hand, and from the typical mass per unit area of a lipid bilayer \cite{Rawicz:2000aa} on the other hand.

The coefficient $\tilde{\lambda}>0$ has properties similar to those of $\lambda$ but only exists if 
the system has a broken-symmetry variable. If the system under study 
were a crystal, we would interpret this coefficient as related to 
the phenomenon of vacancy diffusion, i.e., the displacement of mass 
without change in the periodic lattice. In our system, unlike in a crystal, there can be two independent 
diffusion coefficients $\propto \lambda$ and $\propto \tilde{\lambda}$ even if the creation 
of vacancies (and therefore possibly the breaking of the helix) is 
forbidden. Indeed, one does not need to create holes in the helix 
to displace mass without disturbing the periodic lattice 
of Fig.~\ref{fig:dynamin_symmetry}(b): this can be done by changing 
the radius of the helix. As far as orders of magnitude are concerned, 
we only assume in the following that the transport phenomena associated 
with $\tilde{\lambda}$ are not much faster than the ones associated 
with $\lambda$.

In the following, we consider the system as isothermal. This condition can be enforced by making the thermal conductivity $\kappa>0$ go to infinity, which implies that any thermal gradient relaxes instantaneously. In this $\kappa\rightarrow\infty$ limit, we can drop
Eq.~(\ref{eq:constitutive_1_5}) as well as the last terms of
Eqs.~(\ref{eq:constitutive_1_3}) and~(\ref{eq:constitutive_1_4}).

Eventually, $b$, $c$ and $d$ describe couplings between the three diffusion phenomena described above. Such cross-effects give rise for instance to the so-called Soret and Dufour effects. As in the case of $a$, the positivity of the matrix of phenomenological coefficients sets upper bounds on their values.

\subsection{\label{sec:Coupling of GTP hydrolysis or binding to the dynamics}Coupling of GTP hydrolysis or binding to the dynamics}
We have now developed a complete formalism for the dynamics of a passive, non-polar, diphasic helix submitted to external friction. However, the system considered in this article is not passive since nucleotide (i.e. GTP or GTP analogue in this context) hydrolysis by dynamin or at least binding to dynamin is required for conformational change. In the following, we introduce this external free energy source using arguments similar to those of Ref.~\cite{Kruse:2005aa}. Instead of deriving a whole new formalism taking into account the conservation of GTP, GDP and P$_\text{i}$ and all the chemical reactions involving them, we model the presence of GTP in the experimental chamber by a spatially homogeneous ``chemical force'' $\Delta\mu$, where $\Delta\mu$ stands for the free energy provided by the hydrolysis (or, arguably, binding) of one GTP molecule.

From Table~\ref{tab:fluxes_and_forces}, we see that the spatial symmetry of
$\Delta\mu$ only allows it to couple to Eqs.~(\ref{eq:constitutive_1_1}) 
and (\ref{eq:constitutive_1_2}). We also note that time-reversal 
symmetry imposes that these couplings are reactive. Neglecting 
thermal diffusion as discussed in Sec.~\ref{sec:Discussion of the phenomenological coefficients}, we obtain a modified set of constitutive equations:
\begin{subequations}\label{eq:constitutive_2}
\begin{eqnarray}
\sigma-\mathfrak{p}-h/p &=& -\eta_z \nabla v - a\nabla\Omega+\xi_z\Delta\mu,\label{eq:constitutive_2_1}\\
\tau+h &=& -a \nabla v - \eta_\theta\nabla\Omega+\xi_\theta\Delta\mu,\label{eq:constitutive_2_2}
\end{eqnarray}
\begin{eqnarray}
J &=& - \lambda\nabla\mu_e - b\nabla h,\label{eq:constitutive_2_3}\\
X+\frac{J}{\rho\Phi p} &=& - b\nabla\mu_e - \tilde{\lambda}\nabla h.\label{eq:constitutive_2_4}
\end{eqnarray}
\end{subequations}

\subsection{\label{sec:Hydrodynamic modes}Hydrodynamic modes}
The hydrodynamic relaxation modes are studied by linearizing the equations 
of motion around the state of thermal equilibrium. By definition, all thermodynamic forces vanish at thermal equilibrium, and in particular $\Delta\mu=0$. Let $\delta\rho=\rho-\rho_0$ and $\delta\Phi=\Phi-\Phi_0$ be the deviations of the mass density and of the mass fraction of fluid $h$ from this state. Combining the conservation equations of Sec.~\ref{sec:Conservation laws and hydrodynamic variables} with the constitutive equations Eqs.~(\ref{eq:constitutive_2}) yields dynamical equations relating $\delta\rho$, $\uzt$ and $\delta\Phi$ with $g$, $l$ and the reactive (equilibrium) forces $\mathfrak{p}$, $h$ and $\mu_e$. In the overdamped regime, $g$ and $l$ can be eliminated according to Eq.~(\ref{eq:gl_overdamped}). Close to equilibrium, the reactive forces are linearly related to the state vector of the system $\bm{x}=(\delta\rho, \uzt, \delta\Phi)$ through a susceptibility matrix $\chi$:
\begin{equation}\label{eq:chi_definition}
\left(\begin{array}{c}\mathfrak{p}\\h\\\mu_e\end{array}\right)
=\chi
\left(\begin{array}{c}\delta\rho\\\uzt\\\delta\Phi\end{array}\right)
=\chi\bm{x}.
\end{equation}
It is therefore possible to write linearized dynamical equations for $\bm{x}$ in a closed form. In Fourier space and to leading order in the wave vector $q$, it reads
\begin{equation}\label{eq:hydrodynamic_equation}
i\omega
\bm{x}
=-q^2
\left(A^r\gamma^{-1}B^r+\tilde{A}^d\right)\chi
\bm{x},
\end{equation}
where $A^r$, $B^r$, $\gamma$ and $\tilde{A}^d$ are matrices. $A^r$ and $B^r$ describe reactive couplings, the elements of $\gamma$ are the $\gamma_{ij}$s defined above and $\tilde{A}^d$ contains dissipative phenomenological coefficients. See Appendix~\ref{sec:Derivation of the hydrodynamic modes and perturbation theory} for details. According to Eq.~(\ref{eq:hydrodynamic_equation}), the system has three diffusive hydrodynamic modes.

\section{\label{sec:Long times dynamics for the helix/membrane 
friction-limited regime}Long times dynamics for the 
helix/membrane friction-limited regime}

The results of Section~\ref{sec:Hydrodynamic theory} allow a full description
of the hydrodynamic behavior of the dynamin/membrane tube. For instance, 
to predict the relaxation of a helix with some known initial and 
boundary conditions on the hydrodynamic variables
($\delta\rho,\uzt,\delta\Phi$), one should diagonalize the matrix
\begin{equation}\label{eq:hydrodynamic_matrix}
M=\left(A^r\gamma^{-1}B^r+\tilde{A}^d\right)\chi
\end{equation}
and solve three diffusion equations along the directions defined by 
its eigenvectors. We denote the eigenvalues of $M$ as $D_1>D_2>D_3$. Unfortunately, this diagonalization 
yields very lengthy expressions from which no intuitive picture 
of the dynamics can be deduced. Nevertheless, we show in this section 
that all experimentally observable features of the tube's dynamics 
can be faithfully described by more convenient simplified equations.

The matrix $M$ is the sum of two terms. This reflects the fact that the dynamics of the tube involves two sources of damping: $A^r\gamma^{-1}B^r\chi$ describes the friction against the outer water and $\tilde{A}^d\chi$ is associated with dissipation mechanisms internal to the tube, such as helix/membrane friction. 
We now make an estimate of the orders of magnitude of these two effects.
Assuming that the friction of the tube against 
water is that of a rigid rod of radius $r_e$ (defined as the external radius of the dynamin coat) and length $L$ yields \cite{Guyon:2001aa}
\begin{equation}\label{eq:gamman}
\gamma_{zz}\simeq\frac{2\pi\eta\rho^{-1}}{\text{ln}\left(\frac{L}{r_e}
\right)-0.72},
~
\gamma_{z\theta}=\gamma_{\theta z}=0,
~
\gamma_{\theta\theta}\simeq\frac{4\pi\eta r_e^2}{I},
\end{equation}
where $\eta$ is the viscosity of water \footnote{Note that as far as Eq.~(\ref{eq:simplified_dynamics}) is concerned, the exact values of 
the coefficients of $\gamma^{-1}$ do not matter as long as they are large. Therefore, there would be no point in trying to improve Eq.~(\ref{eq:gamman}).}. We evaluate the coefficients of $A^r\gamma^{-1}B^r$ from these expressions. Section~\ref{sec:Discussion of 
the phenomenological coefficients} provide a similar estimate of $\tilde{A}^d$ (which is consistent with experiments as shown in Sec.~\ref{sec:Time scales}), and the coefficients of $A^r\gamma^{-1}B^r$ are found to be at least four orders of magnitude larger than those of $\tilde{A}^d$. We can therefore diagonalize $M$ perturbatively in $\tilde{A}^d\chi$. Since $A^r\gamma^{-1}B^r\chi$ has one vanishing eigenvalue (see Appendix~\ref{sec:Derivation of the hydrodynamic modes and perturbation theory}), $D_3$ is much smaller than $D_1$ and $D_2$. We write it and the associated eigenvector to lowest order in $\tilde{A}^d\chi$:
\begin{equation}\label{eq:slow_mode}
D_3=\frac{\lambda}{\rho}\frac{\det(\chi)}{
\left\vert
\begin{array}{cc}
\chi_{1,1} & \chi_{1,2} \\
\chi_{2,1} & \chi_{2,2} \\
\end{array}
\right\vert
},~
\bm{x}_3=
\chi^{-1}
\left(\begin{array}{c}0\\0\\1\end{array}\right),
\end{equation}
where the denominator of the second term in $D_3$ is the (3,3) cofactor 
of matrix $\chi$. 

This slow mode can be interpreted as follows. The orders of 
magnitude calculations given above show that $\gamma^{-1}$ is large, 
meaning that the friction of water against the tube is very weak. 
Therefore, according to Eqs.~(\ref{eq:gl_overdamped}) the tube 
quickly (although in hydrodynamic times $\propto q^{-2}$) relaxes 
to a state of constant tension and torque $\nabla \sigma=0$, 
$\nabla\tau=0$. Anticipating on the results of Sec.~\ref{sec:Time scales}, we estimate that this regime is reached in a few tens of milliseconds for a tube 
of $10\,\mu$m. In experimental situations close to that 
of Ref.~\cite{Roux:2006aa}, this relaxation is much faster even than 
the injection of GTP in the experimental chamber. Therefore, 
when interested in observable time scales, one should consider that 
the two fast modes of Eq.~(\ref{eq:hydrodynamic_equation}) are always 
at equilibrium. Using  Eqs.~(\ref{eq:constitutive_2_1}),
(\ref{eq:constitutive_2_2}), we deduce that $\mathfrak{p}$ and $h$ 
have the following spatially homogeneous values:
\begin{subequations}\label{eq:equilibrium_ph}
\begin{eqnarray}
\mathfrak{p} & = & \sigma+\frac{\tau}{p}-\left(\xi_z+
\frac{\xi_\theta}{p}\right)\Delta\mu,\\
h & = & -\tau+\xi_\theta\Delta\mu,
\end{eqnarray}
\end{subequations}
where $\sigma$ and $\tau$ are independent of $z$ and are fixed by 
the boundary conditions imposed on the tube. Note that introducing GTP 
in the system, thus changing the value of $\Delta\mu$, is equivalent 
to applying a force $-\xi_z\Delta\mu$ and a torque $-\xi_\theta\Delta\mu$ 
to the tube.

If the two fast modes are considered at equilibrium, the tube dynamics can be described by the evolution equation of the projection of the state of the system onto the third mode. In our approximation, this projection is $\delta\Phi/\left(\chi^{-1}\right)_{3,3}$. The equations of motion of the system therefore reduce to a single diffusion equation whose diffusion coefficient is the smallest eigenvalue of $M$:
\begin{equation}\label{eq:simplified_dynamics}
\dt \delta\Phi=D_3 \nabla^2 \delta\Phi.
\end{equation}

\section{\label{sec:Susceptibility matrices describing 
experimental situations}Susceptibility matrices describing 
experimental situations}

Although this is already true in the general case 
of Eq.~(\ref{eq:hydrodynamic_equation}), it appears even more clearly
in the simplified Eq.~(\ref{eq:slow_mode}) that a full understanding of 
the dynamics requires an expression of the susceptibility matrix $\chi$. 
Before proposing such expressions, we would like to comment 
on the nature of the assumptions they imply. 
Unlike in the previous sections, where only well-controlled approximations
based on orders of magnitude and the symmetries of the system are used, the calculation
of $\chi$ requires an explicit expression for the free energy of the tube.
As emphasized in the introduction, such a microscopic description 
is difficult given our limited knowledge of the mechanics of dynamin.
Nevertheless, since the models that
we develop in this section are equilibrium models of the tube, 
all the information about the non-equilibrium behavior of the system is
still being captured by the phenomenological coefficients introduced 
in Section~\ref{sec:Constitutive equations}. This means that we do not 
make any assumptions on the microscopic details of the dissipation mechanisms.

In the following, we first define a microscopic parametrization of the dynamin/membrane tube. Then we propose three equilibrium models of the tube, aimed at describing the experimental situations of Refs.~\cite{Stowell:1999aa,Marks:2001aa} and
\cite{Sweitzer:1998aa,Zhang:2001aa,Danino:2004aa,Chen:2004aa,Mears:2007aa},
where different types of lipids were used as templates for dynamin assembly.

\subsection{Microscopic parametrization}
For the sake of simplicity, let us start by idealizing the geometry. 
We first assume that the membrane is infinitely thin. It is confined 
to a roughly cylindrical shape by the dynamin helix but small deviations 
from this shape are allowed in the following. The energy cost of 
such deformations is fixed by the membrane's stretching and 
bending moduli $k_s$ and  $k_b$. The detailed calculation of the 
membrane's bending energy in the geometry considered here is presented 
in Appendix~\ref{sec:Membrane geometry and bending energy}. We furthermore
consider the helix as an infinitely thin inextensible elastic 
rod with spontaneous curvature and torsion, such that its equilibrium 
shape is a helix of radius $r=r_0$ and pitch $2\pi p=2\pi\alpha r_0$. 
Its elasticity is described as that of a classical spring and is
parametrized by its curvature and torsional rigidities $k_\kappa$ and 
$k_\tau$ \cite{Love:1927aa}. All relevant details are presented 
in Appendix~\ref{sec:Elastic properties of the helix}.

The assumption of inextensibility of the rod forming the helix is the most speculative point of this section. Electron micrographs of dynamin helices treated with the non-hydrolyzable GTP analogue GMP-PCP suggest that the number of dynamin subunits per unit of helix length could change upon GTP binding \cite{Chen:2004aa}. We still use the inextensibility assumption for simplicity and by lack of a satisfactory alternative hypothesis.

In the remainder of this article and unless otherwise stated, we express 
all quantities in units of the helix's spontaneous radius $r_0$, the mass per unit length 
$\rho_0$ and the typical force needed to stretch the helix 
${\cal K}\simeq 2.2\times 10^{-8}\,$N (see 
Appendix~\ref{sec:Elastic properties of the helix}). In these units, 
we define the deviations of the radius and pitch of the helix from 
their spontaneous values by $r=1+\delta r$ and $p=\alpha(1+\delta p)$.

\begin{figure}
\resizebox{8.5cm}{!}{\includegraphics{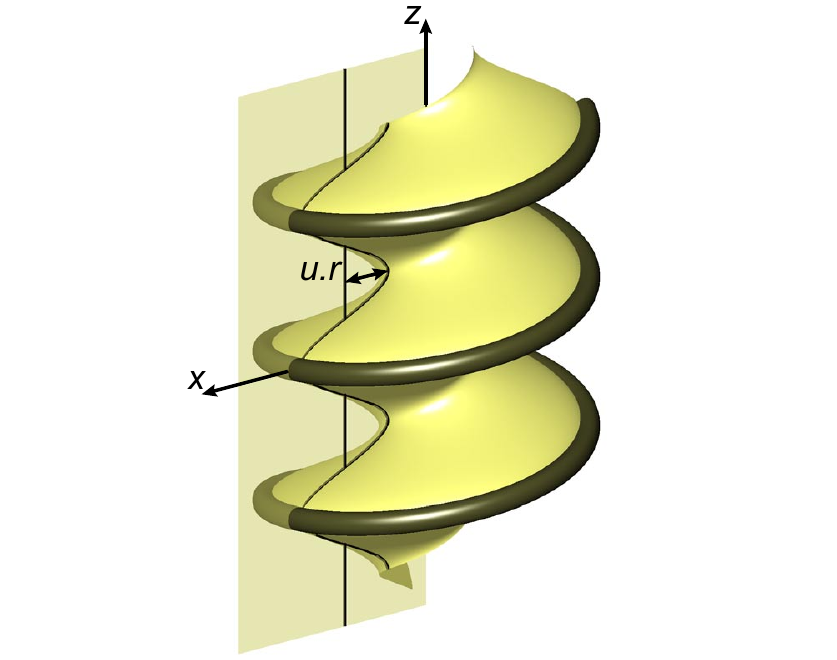}}
\caption{\label{fig:membrane_bending}Sinusoidal deformations of the 
membrane out of its cylindrical shape are allowed in our model. 
The wavy black line materializes the intersection of the membrane 
with the $x>0$, $y=0$ half-plane. The equation of this line is given 
by Eq.~(\ref{eq:wavy_black_line}). Note that it is constrained  to touch 
the helix at each period of the tube.}
\end{figure}

Let $A$ be the area per polar head of lipids and $A_0$ its value 
at equilibrium. We define the relative deviation of $A$ as $a=\frac{A-A_0}{A_0}$.
Eventually, and although the confinement by the protein imposes an 
overall cylindrical shape on the membrane, we allow it to bend 
as long as it retains its helical symmetry. We parametrize this deformation by a number 
$u$ such that the intersection of the membrane with the $x>0$, $y=0$ 
half-plane is the curve (see Fig.~\ref{fig:membrane_bending}):
\begin{equation}\label{eq:wavy_black_line}
x=r(1+u(\cos(pz)-1)).
\end{equation}
Details are given in Appendix~\ref{sec:Membrane geometry and bending energy}.

The elastic properties of the membrane and the helix are such that 
when assembled together, they tend to deform each other: the equilibrium configuration of the tube is different from the spontaneous shapes of the helix and membrane taken separately.
However, we show in the following that these are small effects in the sense 
that at equilibrium, $\delta r\simeq\delta p\simeq a\simeq u\simeq 0$ 
to a good approximation. Therefore, at equilibrium, all the mass of 
the tube is concentrated at a radius $r_0$, hence $I=\rho_0r_0^2=1$.

\subsection{\label{sec:Rigid membrane model}Rigid membrane model}
In the experiments of Refs.~\cite{Stowell:1999aa,Marks:2001aa}, dynamin is assembled on a mixture of non-hydroxylated fatty-acid galactoceramides 
(NFA-GalCer), phosphatidylcholine, cholesterol, 
and phosphatidylinositol-4,5-bisphosphate (PtdIns(4,5)P$_2$). The 
proportions of these lipids are such that even in the absence of dynamin, 
they spontaneously form nanotubes with a diameter comparable to that 
observed for dynamin-coated tubes. Here we investigate 
a suggestion made in Ref.~\cite{Roux:2006aa}, namely that these lipid 
nanotubes are very stiff. Consequently, we model them as rigid 
cylinders ($u=0$) of fixed radius and area per polar head. The last two conditions are imposed by writing the free energy of the tube as
\begin{equation}\label{eq:stiff_energy_1}
f_0=f_h+\frac{k_\infty}{2}\delta r^2+\frac{k'_\infty}{2}a^2,
\end{equation}
where $f_h$ is the elastic energy of the helix, as calculated 
in Appendix~\ref{sec:Elastic properties of the helix}. The assumptions 
$\delta r=0$, $a=0$ are enforced by taking the limit $k_\infty,
k'_\infty\to \infty$. Using the expression Eq.~(\ref{eq:spring_freeenergy}) for $f_h$ and in 
the limit $k_\infty,k'_\infty\to \infty$, one induces no change in 
the dynamics by replacing Eq.~(\ref{eq:stiff_energy_1}) by
\begin{equation}\label{eq:stiff_energy_2}
f_0(\delta r,\delta p, a)=
\frac{k_\infty}{2}\delta r^2+
\frac{k_{pp}}{2}\delta p^2+
\frac{k'_\infty}{2}a^2.
\end{equation}

Since the rod forming the helix is inextensible, its mass density 
is proportional to the rod length per unit length of the tube. The membrane
mass density, on the other hand, is proportional to the radius of 
the membrane cylinder and inversely proportional to the stretching rate of 
the membrane:
\begin{subequations}\label{eq:rhoa_rhob_rm}
\begin{eqnarray}
\rho_h & = & \Phi_0 \frac{\alpha\sqrt{r^2+p^2}}{p\sqrt{1+\alpha^2}}\\
\rho_m & = & (1-\Phi_0)\frac{r}{1+a}.
\end{eqnarray}
\end{subequations}
Combining these equations, one obtains expressions 
for $\delta\rho$ and $\delta\Phi$. One also notices that 
$\uzt=1/p-1/\alpha$, hence the first-order expressions:
\begin{subequations}\label{eq:rm_hydromicro}
\begin{eqnarray}
\delta\rho & = & \left(1-\frac{\alpha^2\Phi_0}{1+\alpha^2}\right)
\delta r-\frac{\Phi_0\delta p}{1+\alpha^2}-(1-\Phi_0)a~~~\\
\uzt & = & -\frac{\delta p}{\alpha}\label{eq:rm_hydromicro_uzt}\\
\delta\Phi & = & \Phi_0(1-\Phi_0)
\left(-\frac{\alpha^2\delta r}{1+\alpha^2}-\frac{\delta p}
{1+\alpha^2}+a\right).
\end{eqnarray}
\end{subequations}

Combining Eqs.~(\ref{eq:stiff_energy_2}) and (\ref{eq:rm_hydromicro}), 
one obtains the function $f_0(\delta\rho,\uzt,\delta\Phi)$. The susceptibility matrix is essentially the matrix of second derivatives of this function (see Eqs.~(\ref{f0s_first_derivatives}) for a more rigorous statement). Taking the limit $k_\infty,k'_\infty\to \infty$, we finally find:
\begin{equation}\label{eq:chi_rm}
\lim_{k_\infty\to \infty\atop k'_\infty\to \infty}\chi_\text{rm}^{-1}
=
\frac{1}{k_{pp}}
\left(
\begin{array}{ccc}
\frac{\Phi_0^2}{(1+\alpha^2)^2} & \frac{\Phi_0}{\alpha(1+\alpha^2)} & \frac{\Phi_0^2(1-\Phi_0)}{(1+\alpha^2)^2}\\
\frac{\Phi_0}{\alpha(1+\alpha^2)} & \frac{1}{\alpha^2} & \frac{\Phi_0(1-\Phi_0)}{\alpha(1+\alpha^2)}\\
\frac{\Phi_0^2(1-\Phi_0)}{(1+\alpha^2)^2} & \frac{\Phi_0(1-\Phi_0)}{\alpha(1+\alpha^2)} & \frac{\Phi_0^2(1-\Phi_0)^2}{(1+\alpha^2)^2}
\end{array}
\right).
\end{equation}

\subsection{\label{sec:Soft membrane models}Soft membrane models}
In many \emph{in vitro} experiments, dynamin is assembled on lipid bilayers containing no cholesterol and which spontaneously form lamellar phases or vesicles in the absence of dynamin \cite{Sweitzer:1998aa,Zhang:2001aa,Danino:2004aa,Chen:2004aa,Mears:2007aa}. From these two observations, we can presume that they are much softer than the lipids studied above and that their spontaneous curvature is zero or at least negligible compared to the curvature imposed by the dynamin coat ($\simeq10^8\,$m$^{-1}$). For these lipids, the microscopic variables $\delta r$, $\delta p$, $a$ and $u$ can therefore all take non-zero values. The free energy of the tube is therefore the sum of three terms: the spring elastic energy of Eq.~(\ref{eq:spring_freeenergy}), a simple quadratic membrane stretching energy with stretching constant $k_s$ and the membrane bending energy of Eq.~(\ref{eq:membrane_energy1}). To second order:
\begin{eqnarray}
f_0&=&
(k_{rr}+2\pi k_b)\frac{\delta r^2}{2}+k_{rp}\delta r\,\delta p+k_{pp}\frac{\delta p^2}{2}
\nonumber\\&&
+2\pi k_s\frac{a^2}{2}+\pi k_b(-\delta r \,u)+k_{uu}\frac{u^2}{2}
\nonumber\\&&
+\pi k_b(-\delta r+u).\label{eq:soft_energy}
\end{eqnarray}
Minimizing this free energy with respect to $\delta r$, $\delta p$ and $u$, we find that at equilibrium
\begin{equation}
\delta r_\text{eq}\sim\delta p_\text{eq}\sim\frac{k_b}{{\cal{K}}r_0}\simeq 2\times 10^{-4}\ll 1
\end{equation}
\begin{equation}
u_{\text{eq}}\sim\frac{k_b}{k_{uu}}\sim\alpha^4\simeq 2\times 10^{-3}\ll 1,
\end{equation}
where we have made use of the fact that $k_{rr},\,k_{rp},\, k_{pp}\sim k_\kappa,\, k_\tau\sim\cal{K}$ in dimensionless units (see Eqs.~(\ref{eq:kkappa_ktau_rod})). We have considered a typical bending modulus $k_b\simeq10\,k_BT$ \cite{Marsh:2006aa} and estimated $\alpha\simeq0.2$ from Ref.~\cite{Danino:2004aa}. These orders of magnitude show that the linear terms in $f_0$ are very small and therefore we neglect the last term of Eq.~(\ref{eq:soft_energy}) in the following. In other words, we use the approximation that at equilibrium the spring assumes its spontaneous radius and pitch $r_\text{eq}=1$, $p_\text{eq}=\alpha$ and that the membrane is an unstretched cylinder of radius $1$ (since $a_\text{eq}=0$ from Eq.~(\ref{eq:soft_energy}) and $u_\text{eq}=0$).

As above, we want to express $f_0$ (now a function of the microscopic variables $\delta r$, $\delta p$, $a$ and $u$) as a function of the hydrodynamic variables $\delta\rho$, $\uzt$, $\delta\Phi$. Since there are four microscopic and three hydrodynamic variables, finding a unique relationship between the two sets of variables seems impossible at first sight. However, there exist constraints on the microscopic variables that we have not yet been expressed.
To understand these constraints, let us calculate two quantities to first order with the help of Eqs.~(\ref{eq:membrane_surface1}) and (\ref{eq:membrane_volume1}): the mass density of lipids, which is the ratio of the surface area covered by the lipids to their area per unit mass:
\begin{equation}\label{eq:rho_lipids}
\rho_\text{lipids}\propto\frac{\mathfrak{s}}{1+a}= 2\pi(1+\delta r-a-u)
\end{equation}
and the volume of water enclosed by the tube:
\begin{equation}\label{eq:inner_water_volume}
\mathfrak{v}=\pi(1+2\delta r-2u).
\end{equation}
If all four microscopic variables were independent, $\rho_\text{lipids}$ and $\mathfrak{v}$ would be independent as well. However, since the membrane tube is filled with water, allowing for a change of $\mathfrak{v}$ at constant $\rho_\text{lipids}$ implies a flow of the water inside the tube relative to the lipids. We estimate the typical time scale associated to this flow to be that of a Poiseuille flow inside a tube of radius $r_0$ driven by a pressure difference ${\cal K}/r_0^2$ and over a distance $L=10\,\mu$m:
\begin{equation}\label{eq:Poiseuille}
t_\text{Poiseuille}= \frac{8\eta(\pi r_0^2L)}{\pi r_0^4}\frac{L}{{\cal K}/r_0^2}\simeq 40\,\mu\text{s}.
\end{equation}
Therefore, on time scales $t\ll t_\text{Poiseuille}$, the relative flow of membrane and inner water is insignificant. Consequently, the ratio of mass density of membrane to mass density of inner water $\rho_\text{lipids}/(\mathfrak{v}\rho_{\text{H}_2\text{O}})$ has to be a constant. Conversely, on time scales $t\gg t_\text{Poiseuille}$, one can consider that the flow of water inside the tube has relaxed, hence $\rho_\text{lipids}$ and $\mathfrak{v}$ are independent variables. On time scales $t\sim  t_\text{Poiseuille}$, the situation is more complex and a correct hydrodynamic theory would involve not two, but three different fluids: the helix, the membrane and the inner water. Such a treatment would obviously be quite heavy and relevant only on experimentally unobservable time scales. In the following, we therefore only calculate $\chi$ in the two limiting cases $t\ll t_\text{Poiseuille}$ and $t\gg t_\text{Poiseuille}$.


\subsubsection{\label{sec:Short time scales: tll ttextPoiseuille}Short time scales: $t\ll t_\text{Poiseuille}$}
In this limit, no relative flow of membrane and inner water is possible and the ratio $\rho_\text{lipids}/(\mathfrak{v}\rho_{\text{H}_2\text{O}})$ is a constant ($\rho_{\text{H}_2\text{O}}=10^3\,$kg.m$^{-3}$ --~the mass per unit volume of water~-- is considered a constant). Using Eqs.~(\ref{eq:rho_lipids}) and (\ref{eq:inner_water_volume}), this yields to first order:
\begin{equation}\label{eq:no_sliding_constraint}
\delta r+a-u=0.
\end{equation}
On top of this constraint, one can write three equations relating the microscopic variables to the hydrodynamic variables. Since the inner water and membrane cannot flow relative to each other, we treat them as a single fluid, which we label ``fluid $m$'', hence $\rho_m=\rho_\text{lipids}+\mathfrak{v}\rho_{\text{H}_2\text{O}}$. Similarly to Eqs.~(\ref{eq:rhoa_rhob_rm}) and using Eqs.~(\ref{eq:rho_lipids}), (\ref{eq:inner_water_volume}) and (\ref{eq:no_sliding_constraint}), one can write to first order:
\begin{subequations}\label{eq:rhoa_rhob_sm}
\begin{eqnarray}
\rho_h & = & \Phi_0 \frac{\alpha\sqrt{r^2+p^2}}{p\sqrt{1+\alpha^2}}\\
\rho_m & = & (1-\Phi_0)(1+\delta r-a-u).\label{eq:rhom_sm}
\end{eqnarray}
\end{subequations}
Moreover, one still has $\uzt=1/p-1/\alpha$, hence up to first order
\begin{subequations}\label{eq:sm_hydromicro}
\begin{eqnarray}
\delta\rho & = & \left(1-\frac{\alpha^2\Phi_0}{1+\alpha^2}\right)\delta r-\frac{\Phi_0\delta p}{1+\alpha^2}\nonumber\\
 & &-(1-\Phi_0)a-(1-\Phi_0)u\\
\uzt & = & -\frac{\delta p}{\alpha}\\
\delta\Phi & = & \Phi_0(1-\Phi_0)\left(-\frac{\alpha^2\delta r}{1+\alpha^2}-\frac{\delta p}{1+\alpha^2}+a+u\right).~~~~~~
\end{eqnarray}
\end{subequations}
Combining these and Eq.~(\ref{eq:no_sliding_constraint}) yields a unique relation between the microscopic and hydrodynamic variables. We differentiate the free energy of Eq.~(\ref{eq:soft_energy}) as a function of the latter, yielding an expression for $\chi_\text{sm}^{t\ll}$. More details are given in Appendix~\ref{sec:Susceptibility matrices}.

\subsubsection{\label{sec:Long time scales: tgg ttextPoiseuille}Long time scales: $t\gg t_\text{Poiseuille}$}
In this limit, the flow of water inside the tube has already relaxed and therefore $\rho_\text{lipids}$ and $\mathfrak{v}$ are independent variables. Consistently with the hydrodynamic approach used in this paper, we consider that the microscopic state of the system has the lowest free energy compatible with the values of the hydrodynamic variables. This yields the following constraint:
\begin{equation}\label{eq:minimization_constraint}
\left.\frac{\partial f}{\partial u}\right|_{\delta\rho,\,\uzt,\,\delta\Phi}=0
\Leftrightarrow
u=\frac{\pi k_b}{k_{uu}}\delta r+\frac{2\pi k_s}{k_{uu}}a.
\end{equation}
As above, this constraint yields a unique relation between the hydrodynamic and microscopic variables. Fluid $m$ now represents only the membrane: $\rho_m=\rho_\text{lipids}$. However Eqs.~(\ref{eq:rhom_sm}) and (\ref{eq:sm_hydromicro}) remain valid. As above, $\chi_\text{sm}^{t\gg}$ is obtained by combining them with the constraint Eq.~(\ref{eq:minimization_constraint}) and the second derivatives of $f_0$. See Appendix~\ref{sec:Susceptibility matrices} for more details.

\section{\label{sec:Comparison to experimental results}
Comparison to experimental results}
In this section we use electron microscopy data to evaluate the active force and torque generated by the tube when supplied with GTP. Using these results, we show that the change of conformation of dynamin is expected to depend strongly on whether it is assembled on a soft or rigid membrane tube, which could account for seemingly contradictory experimental results. We then turn to the tube dynamics and show that although currently available experimental data do not allow a detailed comparison with our theory, the time scales involved are in agreement with our predictions.

The numerical estimates of this section are based on the typical values $k_b\simeq10\,k_BT\simeq4\times 10^{-20}\,$J \cite{Marsh:2006aa} and $k_s\simeq0.25\,$N.m$^{-1}$ \cite{Rawicz:2000aa}. $\eta_\text{water}=9\times 10^{-4}\,$Pa.s, and measurements show $r_0\simeq10\,$nm, $\alpha\simeq0.2$, $r_e\simeq25\,$nm \cite{Danino:2004aa} and ${\cal K}\simeq2.2\times 10^{-8}\,$N (see Sec.~\ref{sec:Persistence length of a helix and experimental determination of cal K}). The water friction and helix elastic constants are calculated from Eqs.~(\ref{eq:gamman}) and (\ref{eq:kkappa_ktau_rod}). We also use $\rho_0\simeq7.5\times 10^{-13}\,$kg.m$^{-1}$ and $\Phi_0\simeq0.5$ (see Sec.~\ref{sec:Discussion of the phenomenological coefficients}).

\subsection{\label{sec:Active terms}Active terms}
According to the symmetry arguments developed in Sec.~\ref{sec:Coupling of GTP hydrolysis or binding to the dynamics}, exposing the tube to GTP yields the same deformation as applying a force $-\xi_z\Delta\mu$ and a torque $-\xi_\theta\Delta\mu$ to it. Making an analogy with a spring submitted to a force and torque, we expect a uniform change of radius and pitch for a dynamin helix incubated with GTP for a very long time. Ignoring fluctuations, this is consistent with experimental data \cite{Sweitzer:1998aa,Zhang:2001aa,Chen:2004aa,Danino:2004aa,Mears:2007aa,Stowell:1999aa,Marks:2001aa}.

Let us first turn to Ref.~\cite{Danino:2004aa}, where the GTP analogue GMP-PCP is used. As discussed is Sec.~\ref{sec:Soft membrane models}, one should describe this system with $\chi_\text{sm}^{t\gg}$ on long time scales. In those experiments, the changes of pitch and radius of the dynamin helix are measured to be
\begin{equation}\label{eq:Danino_conformational_change}
\Delta r=\lim_{t\to\infty}\delta r\simeq - 0.5,~\Delta p=\lim_{t\to\infty}\delta p\simeq - 0.31,
\end{equation}
which is compatible with the results of Refs.~\cite{Sweitzer:1998aa,Zhang:2001aa,Chen:2004aa,Mears:2007aa}. Knowing $\chi_\text{sm}^{t\gg}$, $\Delta r$ and $\Delta p$, we deduce the amplitude of the active terms $\xi_z\Delta\mu$ and $\xi_\theta\Delta\mu$, just like the force and torque exerted on a spring can be deduced from its elastic moduli and the amplitude of its deformation. Considering that no external force or torque are exerted on the tube ($\sigma=0$, $\tau=0$) and assuming that the dynamin-covered portions of the tube are in contact with other regions with which they can freely exchange helix or membrane such that $\mu_e=0$, we combine Eqs.~(\ref{eq:chi_definition}) and (\ref{eq:equilibrium_ph}) to find
\begin{equation}\label{eq:finalhydro}
\left(\begin{array}{c}
\Delta \rho\\
\Delta \uzt\\
\Delta \Phi
\end{array}\right)
=
\left(\chi_\text{sm}^{t\gg}\right)^{-1}
\left(\begin{array}{c}
-\xi_z\Delta\mu-\xi_\theta\Delta\mu/\alpha\\
\xi_\theta\Delta\mu\\
0
\end{array}\right).
\end{equation}
Moreover, according to Appendix~\ref{sec:Susceptibility matrices}, the left-hand side of this equation is a known function of $\Delta r$, $\Delta p$ and $\Delta a=\lim_{t\to\infty}a$. We solve Eq.~(\ref{eq:finalhydro}) in $\xi_z\Delta\mu$, $\xi_\theta\Delta\mu$ and $\Delta a$ and obtain
\begin{equation}\label{eq:active_terms_values}
\xi_z\Delta\mu\simeq5.8\times 10^{-9}\,\text{N},~\xi_\theta\Delta\mu\simeq1.3\times 10^{-18}\,\text{N.m},~\Delta a\simeq0.39.
\end{equation}

\subsection{Variability of dynamin's conformation after GTP hydrolysis}
In contrast with the results presented above, Refs.~\cite{Stowell:1999aa,Marks:2001aa} report that the radius of the dynamin helix does not change upon incubation with GTP and that its pitch does not decrease but increases, yielding $\Delta r'\simeq0$, $\Delta p'\simeq0.7$ (in the following, the dashes denote the deformations associated with these references). It is possible to account for those apparently contradictory results without resorting to biochemical arguments (differences in the type of nucleotide or dynamin used...) but only by the mechanical properties of the lipids.

We assume that the equilibrium properties of the tubes used in these experiments are well described by $\chi_\text{rm}$, as discussed in Sec.~\ref{sec:Rigid membrane model}. Assuming that the biochemistry of the tubes considered here is the same as in Refs.~\cite{Sweitzer:1998aa,Zhang:2001aa,Chen:2004aa,Danino:2004aa,Mears:2007aa} implies that the active terms have the values given in Eq.~(\ref{eq:active_terms_values}). Again assuming that $\mu_e=0$, we combine Eqs.~(\ref{eq:chi_definition}) and (\ref{eq:equilibrium_ph}) to find
\begin{equation}
\left(\begin{array}{c}
\Delta \rho'\\
\Delta \uzt'\\
\Delta \Phi'
\end{array}\right)
=
\left(\chi_\text{rm}\right)^{-1}
\left(\begin{array}{c}
-\xi_z\Delta\mu-\xi_\theta\Delta\mu/\alpha\\
\xi_\theta\Delta\mu\\
0
\end{array}\right).
\end{equation}
It is clear from the assumptions of Sec.~\ref{sec:Rigid membrane model} that combining this equation with Eqs.~(\ref{eq:rm_hydromicro}) and (\ref{eq:chi_rm}) yields $\Delta r'=\Delta a'=0$. More interestingly,
\begin{eqnarray}
\Delta p' & = & -\alpha\uzt\nonumber\\
& = & \left(\xi_z+\left(1-\frac{1+\alpha^2}{\Phi_0}\right)\frac{\xi_\theta}{\alpha}\right)\frac{\Phi_0\Delta\mu}{(1+\alpha^2)k_{pp}}\nonumber\\
& = & \Delta p+\frac{2(1-\alpha^2)(k_\kappa-k_\tau)}{4k_\kappa\alpha^2+k_\tau(1-\alpha^2)^2}\Delta r.\label{eq:dp_McMahon}
\end{eqnarray}
Therefore, we predict that the pitch increases ($\Delta p'>0$) if and only if
\begin{equation}
\frac{k_\tau}{k_\kappa}>\frac{2(1-\alpha^2)\Delta r + 4\alpha^2\Delta p}{2(1-\alpha^2)\Delta r-(1-\alpha^2)^2\Delta p}\simeq 1.5.
\end{equation}

This condition is not satisfied by the cylindrical rod model leading to Eq.~(\ref{eq:kkappa_ktau_rod}). One should however temper this result by considering the crudeness of this model, the limited applicability of our small deformation formalism to the high nucleotide concentration experiments considered in this section, as well as the rather large uncertainty on several numerical values used here. We therefore consider Eq.~(\ref{eq:dp_McMahon}) as a proof of principle that a shrinkage of radius on a soft membrane is compatible with an increase of radius on a rigid membrane.

\subsection{\label{sec:Time scales}Time scales}
Turning to the dynamics of the tube as described in Ref.~\cite{Roux:2006aa}, we apply the perturbation scheme of Sec.~\ref{sec:Long times dynamics for the helix/membrane friction-limited regime} to $\chi^{t\gg}_\text{sm}$ using the typical numerical values presented throughout this paper. We furthermore assume that the size and boundary conditions of the system are such that the smallest wave vector allowed is $q_\text{min}=2\pi/(60\,\mu \text{m})$ \cite{Roux:privatecomm}. Eq.~(\ref{eq:hydrodynamic_equation}) implies that the deformations characterized by $q_\text{min}$ dominate the long-time relaxation of each of the three hydrodynamic modes of the system, yielding three relaxation times $\tau_i=2\pi/D_iq^2_\text{min}$, $i=1,2,3$. For simplicity and without loss of generality, we discuss only these deformations in the following. Finally, we assume that one end of the tube is in contact with a reservoir imposing the boundary condition $\mu_e(z=0)=0$. The relaxation time scales are found to be well-separated: $\tau_1\simeq120\,\mu$s, $\tau_2\simeq37\,$ms and $\tau_3\simeq2.6\,$s. This retrospectively validates the perturbative approach of Sec.~\ref{sec:Long times dynamics for the helix/membrane friction-limited regime}.

\begin{figure}
\resizebox{8.5cm}{!}{\includegraphics{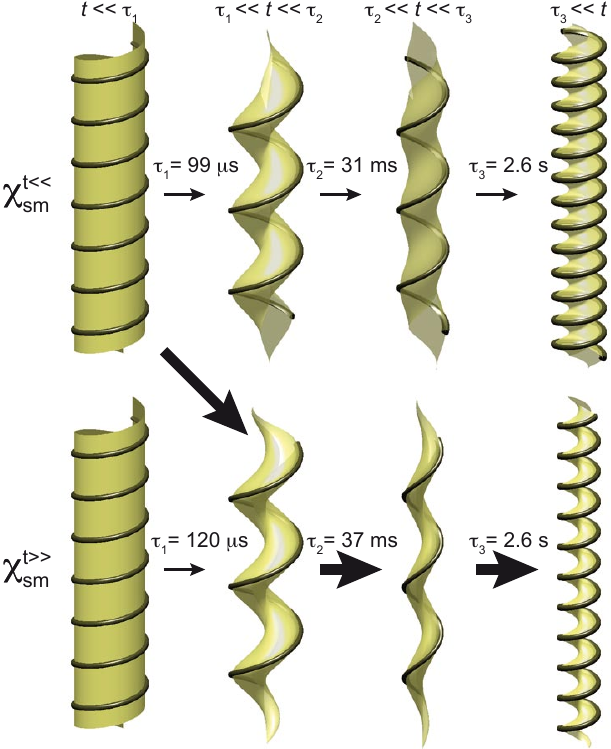}}
\caption{\label{fig:dynamics}Illustration of the dynamics of the tube generated by the susceptibility matrices $\chi^{t\ll}_\text{sm}$ and $\chi^{t\gg}_\text{sm}$. Note that to a good approximation, the field of deformation of the tube is  independent of $z$ during the lag phases between the relaxation of the chronologically well-separated hydrodynamic modes. The amplitude of the changes of conformation are calculated from Eq.~(\ref{eq:active_terms_values}). The transparency of the membrane illustrate its stretching and is proportional to $1+a$. The thick black arrows represent the expected changes of conformation based on the comparison of the $\tau_i$s with $t_\text{Poiseuille}$ (see text).}
\end{figure}

Comparing $\tau_1$ to $t_\text{Poiseuille}$ (Eq.~(\ref{eq:Poiseuille})), we find that $\chi^{t\gg}_\text{sm}$ is probably not a good description of the tube on this time scale and that some intermediate matrix between $\chi^{t\ll}_\text{sm}$ and $\chi^{t\gg}_\text{sm}$ should be used. Looking at Fig.~\ref{fig:dynamics}, however, we realize that the dynamics generated by the two matrices are not very different and that using one or the other does not make much difference at our level of description. On the other hand, it is clear that the transformations characterized by $\tau_2$ and $\tau_3$ must be described using $\chi^{t\gg}_\text{sm}$.

In agreement with Sec.~\ref{sec:Long times dynamics for the helix/membrane friction-limited regime}, $\tau_1$ and $\tau_2$ are smaller than the time needed to inject GTP in the experimental chamber (typically a few tenths of seconds) and are therefore experimentally unobservable: the approximate Eq.~(\ref{eq:simplified_dynamics}) is sufficient to describe current experiments.

On long time scales, our theory predicts an exponential relaxation of the helix with a longest relaxation time $\tau_3\propto 1/(\chi_\text{sm}\lambda q_\text{min}^2)$ (see Eq.~(\ref{eq:slow_mode})). This behavior was indeed reported in Ref.~\cite{Roux:2006aa} and the value $\tau_3=2.6\,$s predicted by us matches the measured value within experimental uncertainty. Note however that our choice of $q_\text{min}$ is to a certain extent arbitrary. Therefore, further measurements are required to obtain a precise value of $\lambda$. Still, even only an order-of-magnitude agreement between the experimental and \emph{a priori} determined value of $\tau_3$ suggests that both the picture of the long-time dynamics of the tube as dominated by the friction between helix and membrane (Sec.~\ref{sec:Discussion of the phenomenological coefficients}) and our description of $\chi_\text{sm}$ are essentially correct.

\section{\label{sec:Discussion}Discussion}
In this article we describe the dynamics of long dynamin-coated membrane nanotubes typically used in \emph{in vitro}, cell-free experiments. This work is therefore relevant to the biological membrane severing function of dynamin insofar as we assume that the tube breaking mechanisms are similar in those two cases. This last section summarizes and discusses our results in this perspective.

Our formalism describes several previously unaccounted for experimental results. Concerning the statics of dynamin, we suggest an explanation for the variability in change of conformation obtained by different experimental groups. Moreover, it is reported in Refs.~\cite{Danino:2004aa,Roux:2006aa} that long tubes incubated with GTP tend to form plectonemic supercoils, which is consistent with our theory. Indeed, in our description, tubes held fixed at both ends and provided with GTP are analogous to rods with persistence length $\ell_p$ under a torque $-\xi_\theta\Delta\mu$ and a compressive force $-\xi_z\Delta\mu$. Therefore, they supercoil if their length is much longer than the critical buckling length $\sqrt{k_BT\ell_p/\xi_z\Delta\mu}\sim\text{a few nm}$, which is always the case in practice. Moving on to the dynamics of the tube, we show that the longest-lived and only experimentally observable internal relaxation phenomenon of the tube is an effective friction mechanism between dynamin helix and lipids. From this we conclude that the internal dynamics of the tube can be approximated by a single diffusion equation, which accounts for the exponential relaxation observed in Ref.~\cite{Roux:2006aa}.

This very simple form for the dynamics of the tube implies robust features that could be tested experimentally: first, the dominant relaxation time should scale like the square of the length of the tube; second, at long times, the rotation frequency of the tube should have a sinusoidal dependence in $z$, which is the shape of the slowest eigenmode of the diffusion equation. Investigating this latter prediction will provide insight into the boundary conditions on the tubes, which could prove useful for making further predictions and may be relevant to tube fission.

In our formalism, nucleotide addition is equivalent to exerting a force and a torque on the tube. Hence, after relaxation of the transient regime, incubation with GTP should induce a homogeneous strain of the dynamin helix. Therefore, two points separated by a distance $z$ should rotate relative to each other a number of times proportional to $z$. This was confirmed by preliminary experimental data \cite{Roux:privatecomm}. This result is intimately linked to the assumption of non-polarity of the dynamin helix, which suggests that the subunits of the dynamin helix themselves are apolar, as proposed in Refs.~\cite{Chen:2004aa,Mears:2007aa}.

We now discuss the assumptions used to derive our formalism. The most important of these is our use of the large-system limit $L\gg r_0$. As discussed in Sec.~\ref{sec:Hydrodynamic theory}, it is obviously correct when applied to the \emph{in vitro}, cell-free experiments considered in this work. Unfortunately, dynamin collars observed in vivo are much shorter~-- typically two to three helical repeats \cite{Koenig:1989aa}. However, we believe that our concepts of friction between helix and membrane and GTP-induced force and torque can be readily transposed to short tubes. One could also be concerned that on small length scales, non-hydrodynamic relaxation phenomena occur on the same time scales as the relaxation phenomena we discuss here and therefore interfere with our picture of the relaxation of the tube. To address this point, let us note that according to Ref.~\cite{Roux:2006aa}, breaking long ($\sim\mu\text{m}$) tubes takes seconds, which is much longer than any reasonable non-hydrodynamic relaxation time for this system. Equivalently, we can say that the tube does not break in short (non-hydrodynamic) times, from which we conclude that non-hydrodynamic internal relaxation phenomena are not essential to tube breaking. Therefore, although the \emph{in vivo} situation is undoubtedly more complex than that considered here, we argue that our description of the internal dynamics of the tube is sufficient to study its tube-severing function.

By writing constitutive equations for the tube, we assumed it to be weakly out of equilibrium. Concerning the friction and viscosity-related phenomenological coefficients (those of Eqs.~(\ref{eq:constitutive_1})), experience shows that this requirement is not very stringent \cite{DeGroot:1984aa}, and our constitutive equations are likely to give a good description of the system in most situations. Chemical systems, on the other hand, typically operate far from equilibrium. In this regime, writing the active forces of Sec.~\ref{sec:Coupling of GTP hydrolysis or binding to the dynamics} as $\Delta\mu$ does not yield the correct dependence on the concentration of GTP. Better results would probably be obtained by using instead $1-e^{-\Delta\mu/k_BT}$, which is characteristic of molecular motors \cite{Julicher:1997aa}. Forthcoming experiments involving low levels of GTP are expected to better fall in the domain of applicability of our theory.

Assumptions of small deviations of the tube from its initial state are also used when deriving the susceptibility matrices $\chi$. Again, these are formally correct at small concentrations of GTP. However, given the fact that those matrices involve uncertain microscopic assumptions, we regard them as nothing more than reasonable examples anyway. Therefore, we do not expect them to yield quantitative results. More reliable information about the characteristics of these matrices could be extracted from micromechanical experiments on dynamin-coated nanotubes.

We now turn to dynamin-induced tube breaking models from the literature. We do not discuss the purely biochemical model of dynamin as a regulatory GTPase --~which has consistently been regarded as unlikely over the past few years \cite{Sever:2000aa,Danino:2001aa,Praefcke:2004aa}~-- and rather concentrate on the mechanochemical models. Depending on authors, the critical feature leading to tube breaking by dynamin has alternately been proposed to be a change of radius \cite{Hinshaw:1995aa}, of pitch \cite{Stowell:1999aa}, or membrane bending \cite{Kozlov:1999aa}. As can be seen from Sec.~\ref{sec:Soft membrane models}, all of these deformations fit very naturally in our theoretical framework. Moreover, we showed in Eq.~(\ref{eq:active_terms_values}) that the change of conformation of dynamin typically induces a significant stretching rate $\Delta a$ of the membrane. We would therefore like to attract attention on this fourth type of deformation of the tube, which might play an important role in tube breaking.

Unlike previous models, this work does not rely on detailed assumptions about the tube's conformational changes. Instead, we predict them by optimally exploiting the experimental data in the light of thermodynamic considerations, conservation laws and symmetry arguments.

Several models have also been proposed for the coupling of GTP to dynamin activity: GTP hydrolysis could induce a concerted conformational change \cite{Marks:2001aa}, while some results suggest that the crucial step is the binding of GTP to dynamin \cite{Sever:1999aa} and others point to a ratchet-like mechanism for its constriction \cite{Smirnova:1999aa}. Since in our framework the coupling of the energy source to the dynamics is deduced from symmetry considerations only, our theory is equally valid in each of these cases.

In addition to including most features previously discussed in the literature, our formalism yields novel quantitative insight into the mechanism of tube breaking. It would be interesting to further discuss Refs.~\cite{Kozlov:1999aa,Liu:2006aa}, where it is assumed that lipids cannot flow through the dynamin-coated region of the tube, in parallel with Ref.~\cite{Kozlov:2001aa}, where this flow is on the contrary considered instantaneous, in the light of our new knowledge of the helix/membrane friction coefficient $\lambda$.

Furthermore, our hydrodynamic point of view could account for some discrepancies between existing models and experiments. Adopting the classification of Ref.~\cite{Sever:2000aa}, we consider models of the ``Garrote'' class --~where the reduction of the dynamin radius pinches the membrane tube to its breaking~-- and of the ``Rigid helix/Elastic membrane'' class --~where the tube breaks because its walls are brought together by a sudden deformation of the helix. If taken at face value and applied to a long tube, these local models predict a uniform density of tube breaks, since what is expected to happen at the neck of a clathrin-coated endocytosis vesicle should happen at every point of the long helix. Experimentally, however, no breaking is observed in such tubes unless their ends are firmly attached to a fixed substrate \cite{Danino:2004aa}. Moreover, attached tubes are observed to straighten upon GTP injection and then break not at several but at a single point \cite{Roux:2006aa}. This sensitivity to distant boundary conditions and spatially inhomogeneous behavior of the tube motivate our description of long-range interactions mediated by tube elasticity and the $z$-dependence of the tube deformation, which could account for the existence of a preferred point of breaking.

In the ``Spring'' model \cite{Stowell:1999aa} as well as in Ref.~\cite{Liu:2006aa}, breaking only occurs at the interface between a dynamin-coated and a bare region of the membrane nanotube. We can imagine that in long tubes, such defects in the dynamin coat either appear during polymerization or that the initially homogeneous dynamin coat breaks upon GTP injection. It is however very unlikely \cite{Roux:privatecomm} that the dynamin coats of Ref.~\cite{Roux:2006aa} have systematically exactly one defect, which would account for the fact that they break at most once. Instead, the tube probably often starts with either many or no defects. In the former case, we have to account for the fact that only one of the defects evolves into a full breaking of the tube. In the latter case, we must explain the creation of a defect in the dynamin coat. It is undoubtedly important to consider the space dependence of the stresses in the tube to answer either of these questions. A mechanism of lipid phase separation similar to that of Ref.~\cite{Allain:2004aa} could also help a defect evolve into a full tube break. Indeed, dynamin is known to strongly bind PtdIns(4,5)P$_2$ \cite{Zheng:1996aa}, and could therefore deplete the bare membrane regions in this lipid. Also, depending on the flow of membrane through the dynamin-coated regions of the tube and on the dynamics of their change of conformation, a bare region might be under more or less stress and therefore break more or less easily.

In conclusion, we developed a complete theoretical framework suited for the analysis of the statics and dynamics of long dynamin-coated membrane nanotubes. We make several predictions concerning the space and time dependence of forces, torques, membrane tension, membrane stretching and helix deformations. We hope that our theory will facilitate the interpretation of forthcoming experimental results and help generate and quantitatively test novel hypotheses on the biological mode of action of dynamin.

\acknowledgments{We would like to thank Patricia Bassereau, Gerbrand Koster and Aur\'elien Roux for attracting our attention to the field of dynamin and for stimulating discussions.}

\appendix
\section{\label{sec:Derivation of the hydrodynamic modes and perturbation theory}Derivation of the hydrodynamic modes and perturbation theory}

This appendix contains details about the derivation (Sec.~\ref{sec:Hydrodynamic modes}) and simplification (Sec.~\ref{sec:Long times dynamics for the helix/membrane 
friction-limited regime}) of the hydrodynamic equations for the tube. Starting from the conservation equations Eqs.~(\ref{eq:rho_Phi_conservation}), (\ref{eq:gl_conservation}), (\ref{eq:uzt_conservation}) and  the constitutive equations Eqs.~(\ref{eq:constitutive_2}) we write the linearized equations of motion for the system:
\begin{equation}\label{eq:hydro_A}
\dt
\left(\begin{array}{c}\delta\rho\\\uzt\\\delta\Phi\end{array}\right)
=A^r\nabla
\left(\begin{array}{c}g\\l\end{array}\right)
+A^d\nabla^2
\left(\begin{array}{c}\mathfrak{p}\\h\\\mu_e\end{array}\right),
\end{equation}
\begin{eqnarray}
\dt
\left(\begin{array}{c}g\\l\end{array}\right)
=
&-&\gamma
\left(\begin{array}{c}g\\l\end{array}\right)\nonumber\\
&+&\left(B^r+\gamma
\left(\begin{array}{ccc}
0 & 0 & \frac{\rho}{\Phi}\\
0 & 0 & 0
\end{array}\right)
A^d\right)\nabla
\left(\begin{array}{c}\mathfrak{p}\\h\\\mu_e\end{array}\right)\nonumber\\
&+&B^d\nabla^2\left(\begin{array}{c}g\\l\end{array}\right),\label{eq:hydro_B}
\end{eqnarray}
where the superscripts ``$r$'' and ``$d$'' denote matrices of reactive and dissipative couplings respectively. These matrices read
$$
A^r=
\left(\begin{array}{cc}
-1 & 0\\
-\frac{1}{\rho p} & \frac{1}{I}\\
0 & 0
\end{array}\right),
~
A^d=
\left(\begin{array}{ccc}
0 & 0 & 0\\
0 & \tilde{\lambda} & b\\
0 & \frac{b}{\rho} & \frac{\lambda}{\rho}
\end{array}\right),
~
\gamma=
\left(\begin{array}{cc}
\gamma_{zz} & \gamma_{z\theta} \\
\gamma_{\theta z} & \gamma_{\theta\theta}
\end{array}\right),
$$
$$
B^r=
\left(\begin{array}{ccc}
-1 & -\frac{1}{p} & 0\\
0 & 1 & 0
\end{array}\right),
~
B^d=
\left(\begin{array}{cc}
\frac{\eta_z}{\rho} & \frac{a}{I} \\
\frac{a}{\rho} & \frac{\eta_\theta}{I}
\end{array}\right).
$$

Let us consider Eqs.~(\ref{eq:free_energy_differential}), 
(\ref{eq:F=F0+Ec}) and (\ref{eq:f0}). They imply that
\begin{equation}\label{f0s_first_derivatives}
\mathfrak{p}=
\rho^2\left.\frac{\partial \left(f_0/\rho\right)}{\partial\rho}\right\vert_{\Phi,\uzt},
~
h=
\left.\frac{\partial f_0}{\partial\uzt}\right\vert_{\rho,\Phi},
~
\mu_e=
\frac{1}{\rho}\left.\frac{\partial f_0}{\partial\Phi}\right\vert_{\rho,\uzt},
\end{equation}
where the temperature $T$ is considered constant. The fields $\mathfrak{p}$, $h$, $\mu_e$ are therefore functions of $\delta\rho$, $\uzt$,
and $\delta\Phi$ only. Close to equilibrium this dependence can be linearized, yielding the definition Eq.~(\ref{eq:chi_definition}) of the susceptibility matrix $\chi$. Using this definition, we can now obtain a closed equation for the hydrodynamic variables. In the presence of friction against water (i.e. if $\gamma$ is positive definite), the left-hand side of Eq.~(\ref{eq:hydro_B}) is irrelevant in the hydrodynamic limit, as shown when going from Eqs.~(\ref{eq:gl_conservation}) to Eqs.~(\ref{eq:gl_overdamped}). To leading order in the wave vector $q$, this, together with Eqs.~(\ref{eq:hydro_A}) and (\ref{eq:chi_definition}) yields the hydrodynamic modes equation Eq.~(\ref{eq:hydrodynamic_equation}), with
\begin{equation}
\tilde{A}^d=
\left(\begin{array}{ccc}
0 & -\frac{b}{\Phi} & -\frac{\lambda}{\Phi}\\
0 & \tilde{\lambda}-\frac{b}{\rho\Phi p} & b-\frac{\lambda}{\rho\Phi p}\\
0 & \frac{b}{\rho} & \frac{\lambda}{\rho}
\end{array}\right).
\end{equation}

We now turn to the perturbative diagonalization of matrix $M=\left(A^r\gamma^{-1}B^r+\tilde{A}^d\right)\chi$, a slight generalization of first-order quantum mechanical perturbation theory to non-hermitian matrices \cite{Cohen-Tannoudji:1973aa}. The important point is that the unperturbed matrix $A^r\gamma^{-1}B^r\chi$ has one vanishing eigenvalue $D_3^0=0$. Indeed, the definitions of $A^r$ and $B^r$ imply that
\begin{equation}
A^r\gamma^{-1}B^r
=
\left(
\begin{array}{ccc}
? & ? & 0\\
? & ? & 0\\
0 & 0 & 0
\end{array}
\right),
\end{equation}
where the question marks stand for non-zero coefficients. Clearly, the vector $\bm{x}_3$ of Eq.~(\ref{eq:slow_mode}) is an eigenvector of $M$ associated with $D_3^0$. To lowest order in $\tilde{A}^d\chi$, $D_3$ is given by Eq.~(\ref{eq:slow_mode}) and $\bm{x}_3$ is the associated eigenvector.

\section{\label{sec:Membrane geometry and bending energy}
Membrane geometry and bending energy}
\begin{figure}
\resizebox{8.5cm}{!}{\includegraphics{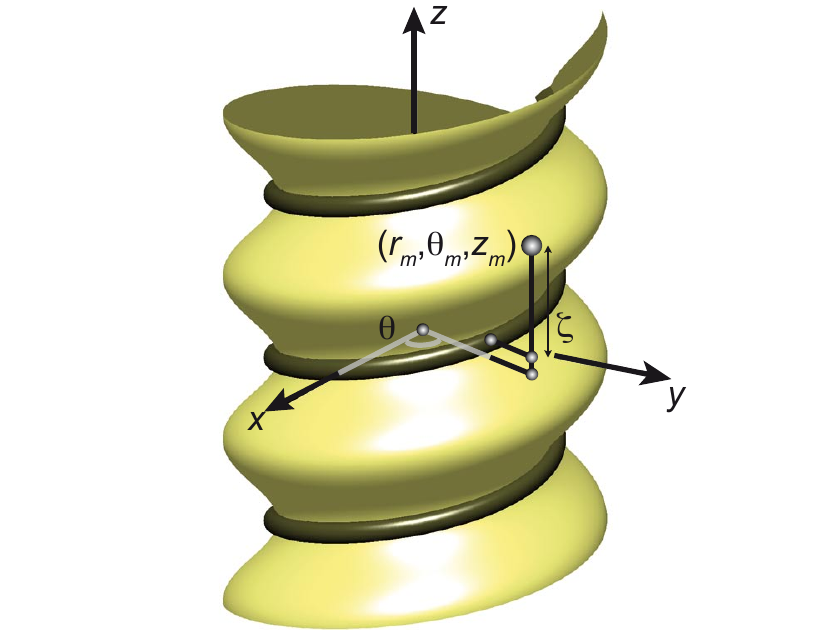}}
\caption{\label{fig:membrane_parametrization}Parametrization of a 
surface confined by a helical scaffold. Any point whose cylindrical coordinates $(r_m, \theta_m, z_m)$ can be written 
as $r_m=r(1+\epsilon(\zeta))$, $\theta_m=\theta$, $z_m=p\theta+\zeta$ 
with $\theta\in [-\infty,\infty]$, $\zeta\in [0,2\pi p]$ belongs 
to the membrane.}
\end{figure}

The calculations of this section are inspired by those 
of Ref.~\cite{Kozlov:1999aa}. In this appendix, we calculate the bending energy of an infinitely thin membrane with no spontaneous curvature surrounded by a helical 
scaffold of radius $r$ and pitch $2\pi p$. We assume that this scaffold 
imposes two constraints on the membrane: first, the membrane has the 
same helical symmetry as the scaffold; and second, the membrane is attached 
to the scaffold and must therefore touch it at every point. 
Under these constraints, the membrane radius as a function of 
the angle $\theta\in [-\infty,\infty]$ of cylindrical coordinates and 
the elevation $\zeta\in [0,2\pi p]$ from the scaffold reads 
(see Fig.~\ref{fig:membrane_parametrization}):
\begin{equation}
r_m(\theta,\zeta)=r(1+\epsilon(\zeta)),
\end{equation}
where $\epsilon(0)=\epsilon(2\pi p)=0$. It can be shown that 
in the $\epsilon\ll 1$ regime, approximating $\epsilon$ by its first Fourier component changes all results presented in this article by less than 1\%. We therefore use this approximation throughout:
\begin{equation}\label{eq:simplified_membrane}
r_m(\theta,\zeta)=r(1+u(\cos(p \zeta)-1))
\end{equation}

The bending free energy of the membrane reads
\begin{equation}\label{eq:bending_free_energy}
F_m=\int\!\!\!\!\int\left(\frac{k_b}{2}{\cal H}^2\right)\,\d \mathfrak{S},
\end{equation}
where $\cal H$ is the total curvature of the membrane and $k_b$ its bending 
modulus. The integration runs over the surface of the membrane. For the configurations considered in this paper, the dependence 
of $k_b$ on the area per polar head of lipids can be 
neglected \cite{Szleifer:1990aa}. We now calculate $F_m$ as a 
function of $r$, $p$ and $\epsilon$ using differential geometry \cite{David:1989aa}. To second order in $\epsilon$, the surface element of the membrane reads
\begin{equation}\label{eq:surface_element}
\d \mathfrak{S}=\left(1+\epsilon+\frac{1}{2}\left(1+\frac{p^2}{r^2}\right)r^2
(\partial_\zeta \epsilon)^2\right)r\,\d\theta\,\d\zeta.
\end{equation}
Integrating this surface element and using Eq.~(\ref{eq:simplified_membrane}), we calculate $\mathfrak{s}$, the membrane area per unit of $z$-length of the tube to first order in $u$:
\begin{equation}\label{eq:membrane_surface1}
\mathfrak{s}=2\pi r\left(1-u\right).
\end{equation}
Similarly, to first order in $u$ the volume enclosed by the membrane by unit 
of $z$-length is
\begin{equation}\label{eq:membrane_volume1}
\mathfrak{v}=\pi r^2\left(
1-2u
\right).
\end{equation}
To second order in $\epsilon$, The total curvature of the membrane reads
\begin{eqnarray}
{\cal H} & = & -\frac{1}{r}(1-\epsilon)+\left(1+\frac{p^2}{r^2}\right)
r (\partial_\zeta^2\epsilon)\nonumber\\
& & -\frac{\epsilon^2}{r}-2\frac{p^2}{r}\epsilon(\partial_\zeta^2\epsilon)+
\frac{1}{2}\left(1-\frac{p^2}{r^2}\right)r(\partial_\zeta\epsilon)^2.
\label{eq:membrane_curvature}
\end{eqnarray}
Performing the integration of Eq.~(\ref{eq:bending_free_energy}) using Eqs.~(\ref{eq:simplified_membrane}), (\ref{eq:surface_element}) and (\ref{eq:membrane_curvature}), we find the bending energy per unit of $z$-length of the tube
\begin{eqnarray}\label{eq:membrane_energy1}
f_m&=&\frac{\pi k_b}{r}\left(
1+u+\left(\frac{3}{2}+\frac{3\alpha^{-2}}{2}+\alpha^{-4}\right)\frac{u^2}{2}
\right)\nonumber\\
&=& \frac{1}{r}\left(\pi k_b
(1+u)+k_{uu}\frac{u^2}{2}
\right).
\end{eqnarray}
This last equality defines $k_{uu}$.

\section{\label{sec:Elastic properties of the helix}
Elastic properties of the helix}
We describe the elasticity of the dynamin helix as that of a simple rod with constant spontaneous curvature and torsion \cite{Love:1927aa}. Its elastic energy therefore reads 
\begin{eqnarray}
F_h & = & \int\left(\frac{k_\kappa}{2}(\kappa(\ell)-\kappa_0)^2+
\frac{k_\tau}{2}(\tau(\ell)-\tau_0)^2\right)\,\d \ell\nonumber\\
& = &
\int \left(\frac{\sqrt{r^2+p^2}}{p}\left\lbrace\frac{k_\kappa}{2}\left(\frac{r}{r^2+p^2}-\frac{r_0}{r_0^2+(\alpha r_0)^2}\right)^2\right.\right.\nonumber\\
& & +
\left.\left.\frac{k_\tau}{2}\left(\frac{p}{r^2+p^2}-\frac{\alpha r_0}{r_0^2+
(\alpha r_0)^2}\right)^2\right\rbrace\right)\,\d z,\label{eq:spring_elasticenergy}
\end{eqnarray}
where the first integral is calculated over the curvilinear length of 
the rod and the second one over the $z$-coordinate (see Fig.~\ref{fig:dynamin_symmetry}(a)). In the second expression 
we replace the curvature $\kappa(\ell)$ and torsion $\tau(\ell)$ by their values for a spring of radius $r$ and pitch $2\pi p$. We also choose the spontaneous curvature $\kappa_0$ and torsion $\tau_0$ such that the ground state of the rod is a helical spring of radius $r_0$ and pitch $2\pi \alpha r_0$.

The dynamin helix binds to the membrane through a specific domain, the PH domain. Let $\bf{u}(\ell)$ be the unitary vector field defined on the helix that always points in the direction of the PH domain. Since the PH domain always faces the membrane, $\bf{u}(\ell)$ always faces the inside of the helix. Hence $\bf{u}(\ell)=\bf{N}(\ell)$, the normal to the helix. Consequently, according to Ref.~\cite{Kamien:2002aa}, the twist density of the helix is exactly equal to its torsion, which allows us to write $F_h$ as a function of $\kappa(\ell)$ and $\tau(\ell)$ only.

\subsection{Curvature and torsion coefficients of a rod.}
In order to calculate the curvature and torsion moduli of the rod, 
we consider it as a rod of cross-section $\pi \left(\frac{r_e-r_0}{2}\right)^2$, where $r_e$ is the outer radius of the dynamin coat. We first define the typical force needed to deform the helix
\begin{equation}\label{eq:definition_of_K}
{\cal K}=\frac{\pi E \left(\frac{r_e-r_0}{2}\right)^4}{r_0^2},
\end{equation}
where $E$ is the Young modulus of the rod. Assuming that the Poisson ratio of the rod is $1/2$, its curvature and elastic moduli read \cite{Landau:1986aa}
\begin{subequations}\label{eq:kkappa_ktau_rod}
\begin{eqnarray}
k_\kappa&=&\frac{\pi E \left(\frac{r_e-r_0}{2}\right)^4}{4}=\frac{1}{4}\\
k_\tau&=&\frac{\pi E \left(\frac{r_e-r_0}{2}\right)^4}{6}=\frac{1}{6},
\end{eqnarray}
\end{subequations}
where the equalities on the right are valid if all quantities are expressed in units of $r_0$, $\rho_0$, $\cal K$. We use these scaled units in the rest of this section.

\subsection{Elastic energy of a spring}
Writing $r=1+\delta r$ and $p=\alpha (1+\delta p)$, the integrand of Eq.~(\ref{eq:spring_elasticenergy}) (i.e. the elastic energy per unit $z$-length) reads to second order in $\delta r$, $\delta p$
\begin{eqnarray}\label{eq:spring_freeenergy}
f_h&=&\frac{\alpha}{(1+\alpha^2)^{7/2}}\times\nonumber\\
& &\left(\left(\alpha^{-2}(1-\alpha^2)^2k_\kappa+4k_\tau\right)
\frac{\delta r^2}{2}\right.\nonumber\\
&&+2\left(1-\alpha^2\right)(k_\kappa-k_\tau)\delta r\delta p\nonumber\\
&&+\left.\left(4\alpha^2k_\kappa+(1-\alpha^2)^2k_\tau\right)
\frac{\delta p^2}{2}\right)\nonumber\\
&=& k_{rr}\frac{\delta r^2}{2}+k_{rp}\delta r\,\delta p+k_{pp}\frac{\delta p^2}{2}.
\end{eqnarray}
This equality defines $k_{rr}$, $k_{rp}$ and $k_{pp}$.

\subsection{\label{sec:Persistence length of a helix and experimental 
determination of cal K}Persistence length of a helix and experimental 
determination of $\cal K$}
We now consider the possibility for the central axis of the helix to bend with a radius of curvature ${\cal R}\gg r_0$. For simplicity, we assume that the radius of the helix remains constant under this deformation. The pitch is allowed to vary over one period of the helix to minimize the energy of the system. The persistence length of the helix can be defined by expanding its elastic energy in powers of $\cal R$:
\begin{equation}
F_h({\cal R})-F_h({\cal R}=0)=k_BT\int_0^L \frac{\ell_p}{2{\cal{R}}^2} \,\d z.
\end{equation}
Using Eq.~(\ref{eq:spring_elasticenergy}), this yields:
\begin{equation}\label{eq:helix_persistence_length}
\ell_p = 
\frac{2\alpha\sqrt{1+\alpha^2}}{k_BT}\frac{k_\tau k_\kappa}
{\alpha^2 k_\tau+k_\kappa}=\frac{\alpha\sqrt{1+\alpha^2}}{k_BT(3+2\alpha^2)}.
\end{equation}

In Ref.~\cite{Frost:2008aa}, the persistence length of a nanotube of 
brain polar lipids and phosphatidylinositol-4,5-bisphosphate coated 
with rat brain dynamin is measured to be $\ell_p=37\pm 4\,\mu$m. 
From this and Eq.~(\ref{eq:helix_persistence_length}), we calculate
\begin{equation}
{\cal K}=\frac{3+2\alpha^2}{\alpha\sqrt{1+\alpha^2}}
\frac{k_BT\ell_p}{r_0^2}\simeq2.2\times 10^{-8}\,\text{N}.
\end{equation}
Using Eq.~(\ref{eq:definition_of_K}), we estimate $E\simeq220\,$MPa, a value of the same order of magnitude as the typical Young modulus of proteins
$E=2\,$GPa \cite{Howard:2001aa}.

\section{\label{sec:Susceptibility matrices}Soft membrane susceptibility matrices}
For clarity's sake, we regroup some expressions associated with the models of Sec.~\ref{sec:Soft membrane models} here. In both limits discussed in Sec.~\ref{sec:Soft membrane models} (short and long time), one combines the relations Eqs.~(\ref{eq:sm_hydromicro}) between the three hydrodynamic and four microscopic variables with a constraint (Eqs.~(\ref{eq:no_sliding_constraint}) and (\ref{eq:minimization_constraint}) respectively) to find a unique linear relation between the hydrodynamic and microscopic variables, which we write:
\begin{equation}\label{eq:Q_def}
\left(\begin{array}{c}
\delta\rho\\ \uzt\\ \delta\Phi
\end{array}\right)
=
Q_\text{sm}
\left(\begin{array}{c}
\delta r\\ \delta p\\ a
\end{array}\right).
\end{equation}
In a unit system such that $\rho_0=1$, Eqs.~(\ref{eq:chi_definition}) and (\ref{f0s_first_derivatives}) imply that $\chi$ is the matrix of second derivatives of the free energy of Eq.~(\ref{eq:soft_energy}) as a function of $(\delta\rho, \uzt, \delta\Phi)$. Hence 
\begin{equation}
\chi_\text{sm}=\left(\left(Q_\text{sm}\right)^{-1}\right)^T K_\text{sm}\left(Q_\text{sm}\right)^{-1}.
\end{equation}
The expressions for the matrices implicated in this formula depend on the time scale considered. They read
\begin{equation}
K_\text{sm}^{t\ll}
=
\left(\begin{array}{ccc}
k_{rr}+k_{uu} & k_{rp} & k_{uu}-\pi k_b \\
k_{rp} & k_{pp} & 0\\
k_{uu}-\pi k_b & 0 & 2\pi k_s+k_{uu}
\end{array}\right),
\end{equation}
\begin{equation}
Q_\text{sm}^{t\ll}=
\left(\begin{array}{ccc}
\frac{\Phi_0}{1+\alpha^2} & -\frac{\Phi_0}{1+\alpha^2} & -2(1-\Phi_0)\\
0 & -\frac{1}{\alpha} & 0\\
\frac{\Phi_0(1-\Phi_0)}{1+\alpha^2} & -\frac{\Phi_0(1-\Phi_0)}{1+\alpha^2} & 2\Phi_0(1-\Phi_0)
\end{array}\right),
\end{equation}
\begin{equation}
K_\text{sm}^{t\gg}
=
\left(\begin{array}{ccc}
k_{rr}-2\pi k_b+\frac{(\pi k_b)^2}{k_{uu}} & k_{rp} & 0 \\
k_{rp} & k_{pp} & 0\\
0 & 0 & 2\pi k_s\left(1+\frac{2\pi k_s}{k_{uu}}\right)
\end{array}\right),
\end{equation}
\begin{widetext}
\begin{equation}
Q_\text{sm}^{t\gg}=
\left(\begin{array}{ccc}
1-\frac{\alpha^2\Phi_0}{1+\alpha^2}-\frac{\pi k_b(1-\Phi_0)}{k_{uu}} & -\frac{\Phi_0}{1+\alpha^2} & -(1-\Phi_0)\left(1+\frac{2\pi k_s}{k_{uu}}\right)\\
0 & -\frac{1}{\alpha} & 0\\
\Phi_0(1-\Phi_0)\left(-\frac{\alpha^2}{1+\alpha^2}+\frac{\pi k_b}{k_{uu}}\right) & -\frac{\Phi_0(1-\Phi_0)}{1+\alpha^2} & \Phi_0(1-\Phi_0)\left(1+\frac{2\pi k_s}{k_{uu}}\right)
\end{array}\right).
\end{equation}
Neither $\chi_\text{sm}^{t\ll}$ nor $\chi_\text{sm}^{t\gg}$ have compact explicit expressions.
\end{widetext}

\bibliography{/Users/Martin/Papers/Bibliography}
\end{document}